\documentclass[11pt]{article}

\usepackage[T1]{fontenc}
\usepackage{jheppub}
\usepackage{wasysym}

\usepackage{tabularx}
\usepackage{physics}
\usepackage{tikz}
\usetikzlibrary{calc}
\usepackage{graphicx} 
\usepackage{ae}
\usepackage{xcolor}

\usepackage{amsfonts,amsmath,amsthm,amssymb,amsbsy}
\usepackage{mathtools}

\usepackage{pifont}
\newcommand{\cmark}{\ding{51}}%
\newcommand{\xmark}{\ding{55}}%
\newcommand \widebar [1] {\overline{#1}}

\title{The Geometry of BCFW for ABJM Loop Integrands}

\author[1]{Livia Ferro,}\emailAdd{l.ferro@herts.ac.uk}
\author[1]{Ross Glew,}\emailAdd{r.glew@herts.ac.uk}
\author[1]{Tomasz \L ukowski,}\emailAdd{t.lukowski@herts.ac.uk}
\author[2]{and Jonah Stalknecht}\emailAdd{jonah.stalknecht@matfyz.cuni.cz}

\affiliation[1]{Department of Physics, Astronomy and Mathematics, \\ University of Hertfordshire, \\  Hatfield, Hertfordshire, AL10 9AB, United Kingdom}
\affiliation[2]{Institute for Particle and Nuclear Physics, \\ Charles University, \\ Prague, Czech Republic}

\abstract{
	In this paper we investigate the loop-level geometry of ABJM theory from the perspective of lightcone geometries in dual space. This geometry admits a natural fibration, where one of the loop variables can be naturally interpreted as living in a \emph{fiber} for each fixed point of a lower-loop geometry. When varying the latter, this leads us to the definition of $L$-loop \emph{half-chambers}, defined such that `half' of the $(L+1)$-loop fiber remains unchanged. We provide a full classification of these half-chambers, and demonstrate a surprising bijection between $n$-point $L$-loop half-chambers and $L$-loop Feynman diagrams for a cubic scalar theory with $n/2$ particles. Consequently, the sum over $L$-loop half-chambers that computes the $n$-point ABJM amplitude is in direct correspondence with the sum over $L$-loop Feynman diagrams that computes the $(n/2)$-point amplitude of $\Tr(\phi^3)$ theory. These Feynman diagrams are also realised geometrically in the structure of the loop fibers. Furthermore, we argue that the half-chamber expansion is equivalent to the loop-level BCFW recursion for ABJM, which arises naturally from our geometric construction. Finally, we will illustrate how $L$-loop {\it chambers} emerge as the intersection of two $L$-loop half-chambers, and we provide concrete examples of this construction.}

\begin{document}
	
\maketitle
\pagebreak

\section{Introduction}
Over the past decade, we have learned that scattering amplitudes can be understood as the canonical differential form of certain positive geometries \cite{Arkani-Hamed:2017tmz}. This idea, first realised for planar $\mathcal{N}=4$ SYM through the amplituhedron \cite{Arkani-Hamed:2013jha,Arkani-Hamed:2017vfh,Damgaard:2019ztj,Ferro:2022abq,He:2021llb}, has since been extended to a variety of theories, including $\Tr(\phi^3)$ via the ABHY associahedron \cite{Arkani-Hamed:2017mur} and  ABJM theory via the ABJM amplituhedron \cite{He:2022cup,He:2023rou}. The positive geometry framework now encompasses a wide range of observables, such as correlation functions \cite{Eden:2017fow} and cosmological wavefunctions \cite{Arkani-Hamed:2017fdk,Benincasa:2024leu}, illustrating its broader applicability.

In this work, we explore the loop-level geometry of ABJM theory from the perspective of lightcone geometries in dual space. The complexity of the ABJM amplituhedron grows rapidly with both particle number and loop order, making direct computation of its canonical form intractable beyond the simplest of cases. This motivates the development of new techniques that decompose the amplituhedron into simpler pieces whose canonical forms are easier to determine. The most famous decomposition is given by the BCFW recursion relations, which provide us with an all-loop formula for the canonical form $\Omega\left[\mathcal{O}_k^{(L)}\right]$ of the ABJM amplituhedron. For example, at six-points one-loop the BCFW recursion yields a sum over four on-shell diagrams given diagrammatically by 
\begin{align}
\Omega[\mathcal{O}_3^{(1)}] \! &= \ \raisebox{-10pt}{\includegraphics[scale=1]{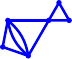}}  \ +\  \raisebox{-10pt}{\includegraphics[scale=1]{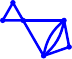}} \ +\ \raisebox{-13pt}{\includegraphics[scale=1]{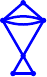}} \ +\  \raisebox{-10pt}{\includegraphics[scale=1]{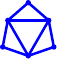}}.
\label{eq:intro_1}
\end{align}
Recently, a complementary {\it fibration} based approach has emerged, naturally formulated in dual momentum space \cite{Lukowski:2023nnf,Ferro:2023qdp,Ferro:2024vwn,Glew:2024zoh}. In its original formulation, the idea behind fibrations is the following. Each point in the tree-level geometry defines a null-polygon ${\bf x}=(x_1,\ldots,x_{2k})$ in dual space, and the lightcones of the cusps $x_i$ bound some compact region $\Delta({\bf x})$. The loop integrand structure of ABJM is then captured geometrically by the space of $L$ points $(y_1,\ldots, y_L)$ inside $ \Delta({\bf x})$ which are additionally required to be mutually spacelike separated. As proposed in \cite{He:2023rou}, by decomposing the tree-level kinematic space into {\it chambers} for which the combinatorial structure of the loop geometry remains constant, the tree-level and loop-level data effectively decouple. This leads to a factorisation of the canonical form into two components: a tree-level piece---the chambers---whose collection tiles the tree-level kinematic space, and a corresponding loop-level factor---the fibers---which encode the loop geometry. In this way, the full geometry can be viewed as a fibration of the loop geometry over the tree-level chambers.

The aim of this paper is to establish a direct connection between the fibration picture and the BCFW recursion for ABJM loop integrands. In contrast to \cite{He:2023rou}, in which the loop geometry is fibrated over tree-level, we consider fibrations of a single loop variable $y_{L+1}$ over the $L$-loop geometry. Specifically, given a point in the $L$-loop amplituhedron, we define the $(L+1)$-loop fiber as the lightcone geometry associated to one additional loop variable $y_{L+1}$. In particular, the vertices of this loop-fiber are located at the intersection of three lightcones, associated to either the cusps of the null-polygon or the preceding loop momenta, and can be classified as either {\it future} or {\it past} vertices. It is then natural to triangulate the $L$-loop amplituhedron into regions for which the set of future or past vertices of the $(L+1)$-loop fiber remain constant. We refer to these regions of the $L$-loop amplituhedron as {\it half-chambers}\footnote{We refer to these as half-chambers as they fix exactly half of the vertices of the fiber.}. 

We find that the $L$-loop half-chambers for $2k$-particle scattering are naturally labelled by $k$-particle, $L$-loop cubic scalar Feynman diagrams, or dually, by triangulations of an $L$-punctured $k$-gon. These diagrams are realised geometrically as the skeletal structure of the $(L+1)$-loop fiber. This is an extension of the results of \cite{Lukowski:2023nnf}, where a similar phenomenon was observed at tree-level. Since the half-chambers tile the full $L$-loop geometry, we arrive at the result that ABJM integrands can be computed by summing over all (tadpole-less) triangulations of an $L$-punctured $k$-gon. An example of the half-chamber expansion of the six-point one-loop integrand is given diagrammatically by 
\begin{align}
	\Omega[\mathcal{O}_3^{(1)}] \! =\raisebox{-17pt}{\includegraphics[scale=0.7]{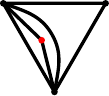}}\ +\ \raisebox{-17pt}{\includegraphics[scale=0.7]{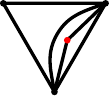}}\ +\ \raisebox{-17pt}{\includegraphics[scale=0.7]{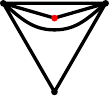}} \ +\ \raisebox{-17pt}{\includegraphics[scale=0.7]{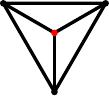}}.
\end{align}
Comparing this to \eqref{eq:intro_1} we see that the terms in the half-chamber expansion are in bijection with the terms in the BCFW expansion. In fact, we will argue that the $L$-loop half-chamber expansion is equivalent to the $L$-loop BCFW recursion for ABJM integrands \cite{Huang:2014xza}. It is remarkable that a simple question about the combinatorial structure of loop fibers naturally gives rise to the BCFW decomposition of the ABJM amplituhedron. Moreover, this correspondence provides a direct geometric interpretation of these recursion relations.

The remainder of this paper is organised as follows. In section \ref{sec:kinematics} we review the basic kinematic facts relevant for studying  ABJM integrands in dual space. In section \ref{sec:ABJM-amp} we detail the construction of the ABJM momentum amplituhedron at tree-level, and the loop-level lightcone geometries. This includes an introduction to the chambers and fibers construction. In section \ref{sec:half-chambers} and \ref{sec:BCFW} we present the main results of this paper, a detailed discussion of half-chambers and their connection to the BCFW recursion relations.  Specifically, we establish that $L$-loop half-chambers have a natural labelling inherited from the structure of the $(L+1)$-loop fiber in terms of triangulations of an $L$-punctured polygon. Moreover, we show a diagrammatic correspondence between the BCFW recursion relations and the $L$-loop Berends-Giele recursion for cubic scalar Feynman diagrams. In section \ref{sec:loop-chambers} we give a brief discussion of full $L$-loop chambers, which arise as the intersection of two $L$-loop half-chambers. Finally, we conclude with a summary and outlook to some open problems in section \ref{sec:conclusions}.

\section{Kinematics}
\label{sec:kinematics}
We will be working in three-dimensional Minkowski space $\mathbb{R}^{1,2}$ with signature $(-,+,+)$ such that the distance between two points $x$ and $y$ is given by
\begin{align}
	(x-y)^2=-(x^1-y^1)^2+(x^2-y^2)^2+(x^3-y^3)^2.
\end{align}
Given a point $x\in\mathbb{R}^{1,2}$ we can split Minkowski space into points which are null separated, positively separated (space-like) and negatively separated (time-like) from $x$ for which we use the notation
\begin{align}
	\mathcal{N}_x &:= \{ y\in \mathbb{R}^{1,2} \ | \ (y-x)^2=0 \}, \notag \\
	\mathcal{N}_x^+ &:= \{ y\in \mathbb{R}^{1,2} \ | \ (y-x)^2>0 \}, \notag \\
	\mathcal{N}_x^- &:= \{ y\in \mathbb{R}^{1,2} \ | \ (y-x)^2<0 \}.
\end{align}  
The scattering data for $n$-point amplitudes in ABJM theory is given by a set of $n=2k$ three-dimensional momenta $p_a^\mu$, $a=1,\ldots,2k$, $\mu=0,1,2$, subject to momentum conservation and the massless on-shell condition $p^2=0$. By convention, we choose all particles with odd/even labels to be outgoing/incoming, respectively, such that momentum conservation is given by 
\begin{align}
	\sum_{a \in \text{odd}}p^\mu_a=\sum_{a \in \text{even}}p^\mu_a.
	\label{eq:dual_coords}
\end{align}
Since we will be dealing with the planar theory, the scattering data can alternatively be encoded using dual momentum coordinates $x_a^\mu$ defined as 
\begin{align}
	p_a^\mu:=x_{a+1}^\mu-x_a^\mu.
\end{align}
Through this relation the scattering data specifies a polygon in Minkowski space with vertices ${\bf x} := (x_1,\ldots,x_{2k})$ whose consecutive vertices $x_a$ and $x_{a+1}$ are null separated. The definition of the dual momentum coordinates is invariant under a global translation and it is convenient to make the choice $x_1=0$ allowing us to invert relation \eqref{eq:dual_coords} to find
\begin{align}\label{eq:xvar-def}
	x_b^\mu = \sum_{a=1}^{b-1}(-1)^a p_a^\mu.
\end{align}
A concept which will play an important role when coming to study the structure of the ABJM momentum amplituhedron is the notion of triple-cut points. Given three points $x_i$, $x_j$ and $x_k$, there generically exist two points $q^\pm_{ijk}$, which we refer to as triple cut points, satisfying the triple-cut conditions
\begin{align}
	(q_{ijk}^\pm-x_i)^2=(q_{ijk}^\pm-x_j)^2=(q_{ijk}^\pm-x_k)^2=0.
\end{align} 
The sign in the superscript of these triple intersections is determined by the orientation of the tetrahedron with corners $(x_i,x_j,x_k,q^\pm_{ijk})$.

As is familiar from its four-dimensional counterpart, the massless on-shell condition, $p^2=0$, can be resolved via the introduction of three-dimensional spinor-helicity variables as
\begin{align}\label{eq:3D-spinor-helicity}
	p^{\alpha \beta}=\left(\begin{matrix} -p^0+p^2 & p^1\\p^1&-p^0-p^2 \end{matrix}\right) = \lambda^\alpha \lambda^\beta.
\end{align}
In the spinor-helicity formalism the following Lorentz invariant brackets appear frequently
\begin{align}
	\langle ab \rangle:= \lambda^1_a \lambda^2_b - \lambda^2_a \lambda^1_b. 
\end{align}

\section{The ABJM  momentum amplituhedron}\label{sec:ABJM-amp}
The tree-level ABJM momentum amplituhedron $\mathcal{O}_k$ is a $(2k-3)$-dimensional geometry in 3D spinor-helicity space which encodes $(n=2k)$-particle scattering in ABJM theory at tree level \cite{He:2021llb,Huang:2021jlh}. A point $\lambda\in\mathcal{O}_k$ is defined to satisfy 
\begin{align}
	&\langle i i+1 \rangle >0,\quad s_{ii+1\ldots j}>0,\\
	&\{\langle 12\rangle, \langle 13 \rangle,\ldots, \langle 1n \rangle \}\quad\text{has $k$ sign-flips},
\end{align}
where we define the planar Mandelstam variables as
\begin{align}
	s_{ii+1\ldots j}\coloneqq \sum_{i\leq a < b\leq j} (-1)^{a+b+1}\langle ab\rangle^2.
\end{align}
Translating into dual space, a point $\lambda\in\mathcal{O}_k$ defines a null-polygon ${\bf x}\coloneqq (x_1^\mu,\ldots, x_n^\mu)$ by making use of equations \eqref{eq:3D-spinor-helicity} and \eqref{eq:xvar-def}. The positivity of the planar Mandelstam variables is now equivalent to the statement that $s_{ii+1\ldots j-1}=(x_i-x_j)^2\geq0\,,\forall i,j$.

\begin{figure}[h]
	\centering
	\begin{minipage}{0.35\textwidth}
		\centering
		\includegraphics[width=\textwidth]{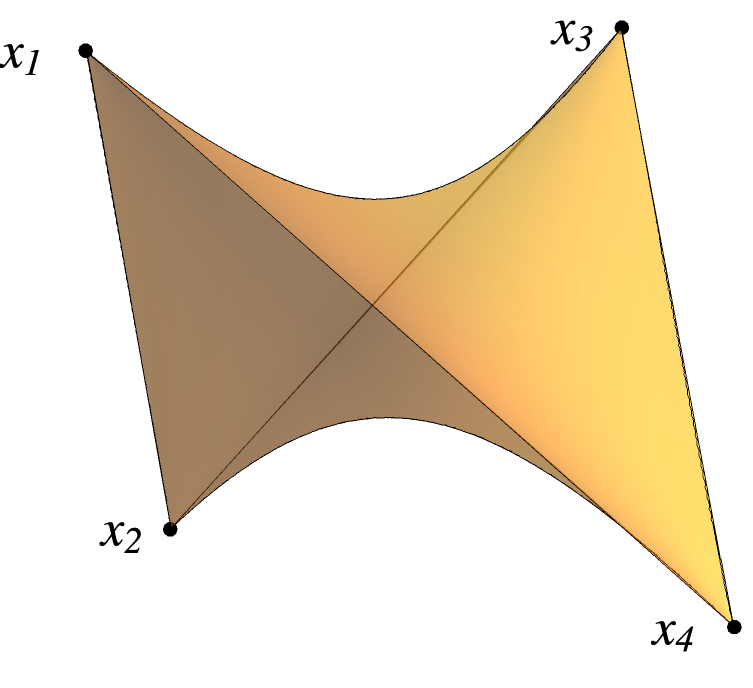} 
		\label{fig:image1}
	\end{minipage}
	\hfill 
	\begin{minipage}{0.55\textwidth}
		\centering
		\includegraphics[width=\textwidth]{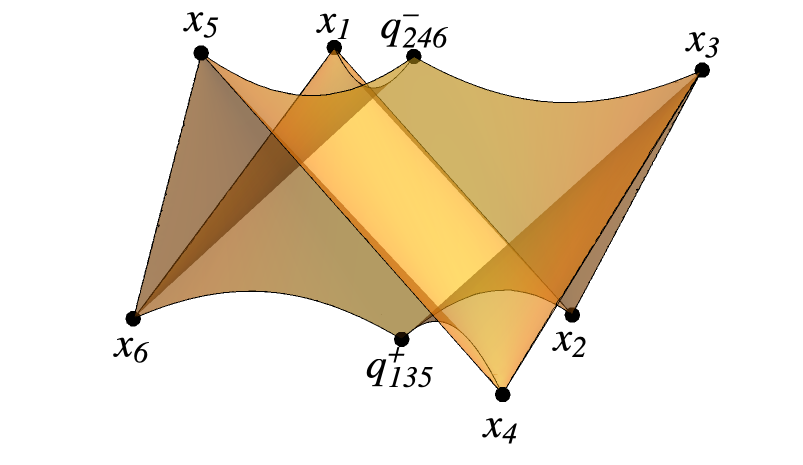} 
		\label{fig:image2}
	\end{minipage}
	\caption{A representative of the one-loop fibers for four and six-points, respectively.}
	\label{fig:one_loop_fibers}
\end{figure}

To generalise the ABJM momentum amplituhedron to loop level, we consider the space of points which are space-like separated from all points $x_i$ in a null-polygon generated by a point in $\mathcal{O}_k$. As observed in \cite{Lukowski:2023nnf} the space $\mathcal{N}_{x_1}^+\cap\ldots\cap\mathcal{N}_{x_n}^+$ splits up into two parts: a non-compact part which extends out to space-like infinity, and a compact part `inside' the null-polygon. The compactness condition is equivalent to the sign-flip condition of \cite{He:2023rou}. We will be interested in this compact part, which we denote by
\begin{align}
	\Delta({\bf x}) &= \{ y \in \mathbb{R}^{1,2}  \ | \ (y-x_i)^2 \geq0, \text{ compact} \}.
\end{align}
We refer to $\Delta({\bf x})$ as the \emph{one-loop fiber}, for reasons that will become clear shortly. Examples of one-loop fibers for $n=4$ and $n=6$ are depicted in Fig. \ref{fig:one_loop_fibers}. 

To extend to all loops, we define
\begin{align}
	\Delta^{(L)}({\bf x}) &= \{ (y_1,y_2,\ldots,y_L) \in (\Delta({\bf x}))^L \ | \ (y_i-y_j)^2\geq0 \quad\forall\; i,j=1,2,\ldots,L \}.
\end{align}
Then the \emph{$L$-loop ABJM momentum amplituhedron} is defined as 
\begin{align}
	\mathcal{O}_k^{(L)} \coloneqq \{(\lambda,\Delta^{(L)}({\bf x}_\lambda)) \ | \ \lambda \in \mathcal{O}_k\},
\end{align}
and its canonical form $\Omega[\mathcal{O}_k^{(L)}]$ computes the $L$-loop integrand in ABJM theory. Here we use the notation ${\bf x}_\lambda$ to indicate the null-polygon which is generated by $\lambda$.

\subsection{Chambers and fibers}
In this section we review the `chambers and fibrations' construction which was recently introduced to study positive geometries. The key idea is to consider a map  which projects out degrees of freedom from a certain geometry. The preimage of this projection map is what is called \emph{fiber}. By studying how the mathematical description of the fibers changes, we are naturally led to a decomposition of the image of this projection into a finite number of \emph{chambers}. The full geometry is then reconstructed as $\bigcup_{i} \mathfrak{c}_i\times f_i$, where the union is over all chambers $\mathfrak{c}_i$ and their associated fibers $f_i$. This framework has been utilised successfully for loop integrands in ABJM in \cite{He:2023rou,Lukowski:2023nnf} and for $\mathcal{N}=4$ SYM in \cite{Ferro:2023qdp,Ferro:2024vwn}, and has been used at tree-level for $\Tr(\phi^3)$ and the inverse KLT kernel in \cite{Bartsch:2025mvy}. See \cite{He:2024xed,He:2025rza} for similar progress for correlation functions.

In our context, at one-loop, the combinatorial structure of the one-loop fiber $\Delta({\bf x})$ can change as we allow the null-polygon ${\bf x}$ to vary. This leads us to the definition of the tree-level chambers as subsets of the tree-level ABJM momentum amplituhedron for which the combinatorial structure of the one-loop fiber $\Delta({\bf x})$ remains constant. Here, the \emph{combinatorial structure} essentially refers to the skeleton of the geometry. That is, we care about which vertices are part of the geometry, how they are connected via edges, and on which facets they lie, but the explicit coordinates of these vertices are not important. An example of this construction is displayed for eight-points in Fig. \ref{fig:tree_chambers_8}. In this case, the tree-level kinematic space is divided into four chambers. The blue lines in each figure encode the one-skeleton of the one-loop fiber. We will return to the full classification of the tree-level chambers in the following section.

To study higher-loop integrands, a number of different approaches has been taken. The most direct is to consider a fibration of the full $3L$-dimensional loop structure over the tree-level \cite{He:2023rou}. This splits the tree-level space into chambers which are further and further refined as the loop level increases. We explain this refinement from the perspective of lightcone geometries in appendix \ref{sec:app_refinement}. In contrast, \cite{Ferro:2024vwn,Glew:2024zoh} preferred an approach where each loop variable is iteratively seen as a fibration over the previous loop variable, creating a stack of `fibrations of fibrations'. While successfully implemented to compute the two-loop MHV momentum amplituhedron form, and the MHV ladder geometries for all $n$, applying this technique naively appears to break down at higher loops/helicity. We comment on this phenomenon more fully in the conclusions. 

In this paper we favour yet another approach, where we fibrate the $(L+1)$\textsuperscript{st} loop variable over the $L$-loop ABJM momentum amplituhedron. More precisely, we define the $(L+1)$-\textit{loop fiber} as the space
\begin{align}\label{eq:loop-fiber-def}
	\Delta({\bf x}; y_1,\ldots,y_L)\coloneqq \{y_{L+1}\in \Delta({\bf x}) \ | \ (y_{L+1}-y_i)^2\geq 0\,,\quad i=1,\ldots, L \}.
\end{align}
Furthermore, we define the $L$-\emph{loop chamber} as a subset of $\mathcal{O}^{(L)}_{k}$ for which the combinatorial structure of the $(L+1)$-loop fiber $\Delta({\bf x}; y_1,\ldots, y_L)$ remains constant. The reason for this choice is to make a close connection to the BCFW recursion relations, which act one loop at a time. We will return to this point in detail in section \ref{sec:BCFW}. 
\subsection{Tree-level chambers and one-loop fibers}\label{sec:tree-level-chambers}
In this section, we present a concise overview of the tree-level chambers and one-loop fibers. The task is now clear: fix a null polygon ${\bf x}$ at tree-level, compute the one-loop fiber geometry $\Delta({\bf x})$, and record the corresponding set of vertices $\mathcal{V}[\Delta({\bf x})]$.  As the tree-level data ${\bf x}$ varies, so too will the vertex set $\mathcal{V}[\Delta({\bf x})]$. A full classification of the tree-level chambers was provided in \cite{Lukowski:2023nnf}, as we now review. 

\begin{figure}[t]
	\centering
	\includegraphics[width=0.22\textwidth]{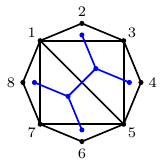}\includegraphics[width=0.22\textwidth]{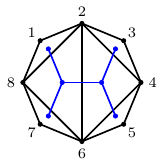} \hfill \includegraphics[width=0.22\textwidth]{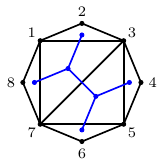}\includegraphics[width=0.22\textwidth]{Figures/eight_point_tree_even_2} \\
	\includegraphics[width=0.22\textwidth]{Figures/eight_point_tree_odd_2}\includegraphics[width=0.22\textwidth]{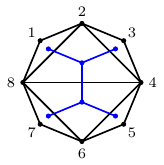} \hfill \includegraphics[width=0.22\textwidth]{Figures/eight_point_tree_odd_1}\includegraphics[width=0.22\textwidth]{Figures/eight_point_tree_even_1}
	\caption{The four tree-level chambers for $n=8$ labelled by a pair of odd and even triangulations of the $n$-gon.}
	\label{fig:tree_chambers_8}
\end{figure}

The set of vertices of any one-loop fiber consists of the cusps of the null polygon $x_i$, together with additional vertices arising from the triple intersections of lightcones $\mathcal{N}_{x_i} \cap \mathcal{N}_{x_j} \cap \mathcal{N}_{x_k}$. These \emph{triple-cut vertices} come either in the type $q^+_{ijk}$ if all $i,j,k$ are odd, or $q^-_{ijk}$ if all $i,j,k$ are even. Remarkably, two odd vertices $q^+_{ijk}$ and $q^+_{lmn}$ can appear together in the one-loop fiber only if, when represented as triangles $t_{ijk}$ and $t_{lmn}$ inscribed on the polygon with odd vertices $(1,3,\ldots,{n-1})$, the two triangles are non intersecting. A similar statement exists for the set of even triple-cut vertices. 

The upshot is that the tree-level chambers can be labelled by a pair $T=(O,E)$ of triangulations of an $n$-gon. The pair consists of one triangulation $O$ made from triangles $t_{ijk}$ containing only odd vertices of the $n$-gon, and $E$ made from triangles containing only even vertices such that
\begin{align}
	t_{ijk}\in O \implies q^+_{ijk} \in \mathcal{V}[\Delta_T({\bf x})],\qquad t_{ijk}\in E \implies q^-_{ijk} \in \mathcal{V}[\Delta_T({\bf x})],
\end{align}
where we have introduced the notation $\Delta_{T}({\bf x})$ for the one-loop fiber associated to $T$. We will refer to these respectively as the \emph{odd}- and \emph{even}-triangulations. Note, the odd (resp. even) triangulations are forced to contain all chords of the form $(ii+2)$ for $i$ odd (resp. even). This combinatorial structure is also known as a \emph{scaffolding triangulation}, and has recently been important in the relation between stringy $\Tr(\phi^3)$ and Yang-Mills amplitudes \cite{Arkani-Hamed:2023swr, Arkani-Hamed:2023jry}.

\begin{figure}[t]
	\centering
	\begin{minipage}{0.5\textwidth}
		\centering
		\includegraphics[width=\textwidth]{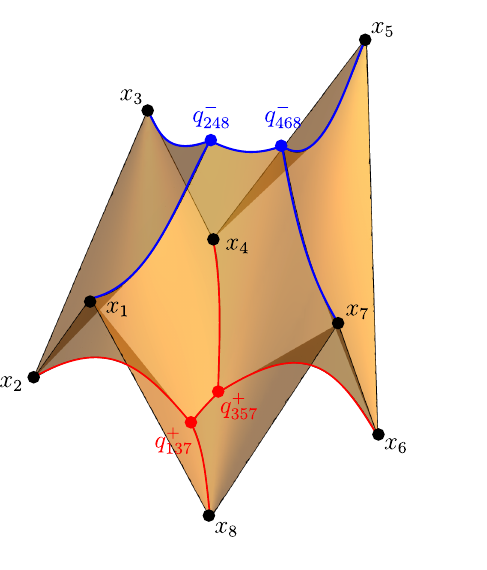} 
		\label{fig:Delta8}
	\end{minipage}
	\hfill
	\begin{minipage}{0.4\textwidth}
		\centering
		\includegraphics[width=\textwidth]{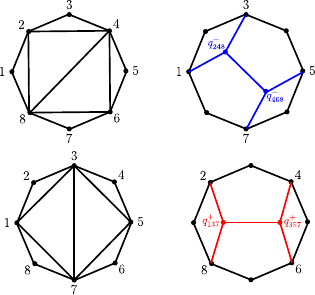}
		\label{fig:Delta8-chamber-label}
	\end{minipage}
	\caption{The one-loop fiber $\Delta_{T_{22}}({\bf x})$ for $n=8$. The corresponding odd/even triangulations $O_2=(t_{137},t_{357}),\,E_2=(t_{248},t_{468})$ are depicted on the right. We see that the dual of these triangulations are tree-level four-point Feynman diagrams which capture part of the structure of $\Delta_{T_{22}}({\bf x})$.}
	\label{fig:8pt_loop_fiber}
\end{figure}

In fact, the \emph{dual} of these triangulations are graphically equivalent to tree-level Feynman diagrams of $\Tr(\phi^3)$ theory, which appear geometrically as the one-skeleton of $\Delta_T({\bf x})$.  As an example at eight points, the odd and even triangulations which label the tree level chambers are given by 
\begin{align}
T_{11}&= (O_1,E_1)= (\{t_{135},t_{157}\}, \{t_{246},t_{268}\}), \ T_{12}=(O_1,E_2)= (\{t_{135},t_{157}\}, \{t_{248},t_{468}\}), \notag \\
T_{21}&=(O_2,E_1)=(\{t_{137},t_{357}\}, \{t_{246},t_{268}\}), \ T_{22}=(O_2,E_2)=(\{t_{137},t_{357}\}, \{t_{248},t_{468}\}).
\label{eq:chams_8_exam}
\end{align}
The dual of these triangulations are displayed in blue in Fig. \ref{fig:tree_chambers_8}, and are equivalent to the one-skeleton of the corresponding one-loop fibers. An explicit example of a one-loop fiber for the triangulation $T_{22}$ is depicted in Fig. \ref{fig:8pt_loop_fiber}.

From the definition of chambers it is clear that the union of all tree-level chambers recovers the full tree-level geometry without overlap, and thus provides a \emph{triangulation} of the positive geometry.  In this way the chamber decomposition expresses the tree-level amplitude as a sum over the Catalan number squared many terms as 
\begin{align}
	\Omega[\mathcal{O}_k] = \sum_{T} \Omega[\mathfrak{c}_{T}],
\end{align}
where the sum runs over all pairs $T=(O,E)$ of odd and even triangulations of the $n$-gon and we have introduced the notation $\mathfrak{c}_{T}$ for the tree-level chamber labelled by $T$ .  

The one-loop ABJM momentum amplituhedron on the other hand can be decomposed as a sum over all products of tree-level chambers times their one-loop fibers as
\begin{align}
	\mathcal{O}_k^{(1)}=\bigcup_{T} \mathfrak{c}_T\times \Delta_T({\bf x}).
\end{align}
This is a relation which extends to the canonical forms of these objects, which allows us to arrive at a formula for the one-loop integrand as
\begin{align}\label{eq:one-loop-chamber-fib}
	\Omega\left[\mathcal{O}_k^{(1)}\right]=\sum_T \Omega[\mathfrak{c}_T]\wedge\Omega[\Delta_T({\bf x})].
\end{align}
Remarkably, the canonical form of the one-loop fiber can be found directly from knowing the vertices of the geometry. The canonical form of the one-loop fiber is immediately found to be
\begin{align}
	\Omega[\Delta_T({\bf x})] = \sum_{t_{ijk}\in O} \omega^+_{ijk} + \sum_{t_{ijk}\in E} \omega^-_{ijk}+\sum_{i=1}^{2k} (-1)^i \omega_{i-1ii+1},
\end{align}
where we define the forms
\begin{align}
	\omega_{ijk}&\coloneqq \dd\log(y-x_i)^2\wedge\dd\log(y-x_j)^2\wedge\dd\log(y-x_k)^2,\\
	\omega^\triangle_{ijk}&\coloneqq \dd\log\frac{(y-x_i)^2}{(y-q^+_{ijk})^2}\wedge\dd\log\frac{(y-x_j)^2}{(y-q^+_{ijk})^2}\wedge\dd\log\frac{(y-x_k)^2}{(y-q^+_{ijk})^2},\\
	\omega^\pm_{ijk}&\coloneqq \omega^\triangle_{ijk} \pm \omega_{ijk}.
\end{align}

\section{Half-chambers}\label{sec:half-chambers}
It is clear from the previous section that the combinatorial structure of the one-loop fiber $\Delta({\bf x})$ is split into an odd side with facets $\mathcal{N}_{x_i}$ for $i$ odd, and an even side with facets $\mathcal{N}_{x_i}$ for $i$ even. This is a consequence of the fact that lightcones naturally split up into a \emph{future} $\uparrow$ and a \emph{past} $\downarrow$ lightcone:
\begin{align}
	\mathcal{N}_x^{\uparrow}\coloneqq \{y\in\mathcal{N}_x \ | \ y^0>x^0\},\qquad \mathcal{N}_x^{\downarrow}\coloneqq \{y\in\mathcal{N}_x \ | \ y^0<x^0\}.
\end{align}
It is natural to keep track of these two pieces of information separately and to examine the structure of the odd and even facets independently. This leads to the intuitive notion which we refer to as odd (resp.~even) \emph{tree-level half-chambers}: regions of $\mathcal{O}_k$ for which the intersection patterns of the odd (resp.~even) lightcones remain unchanged. Again we will label these half-chambers by their respective odd $\mathfrak{c}_O$ and even $\mathfrak{c}_E$ triangulations. We will define $\text{skel}^{\uparrow}({\bf x}_{\text{even}})$ and $\text{skel}^{\downarrow}({\bf x}_{\text{odd}})$ as the combinatorial structure of intersections of $\mathcal{N}^{\uparrow}_{x_2},\ldots,\mathcal{N}^{\uparrow}_{x_{2k}}$, and $\mathcal{N}^{\downarrow}_{x_1},\ldots,\mathcal{N}^{\downarrow}_{x_{2k-1}}$, respectively. The odd and even half-chambers are then defined as\footnote{Here we have introduced the notation ${\bf x}_{\text{even}}=x_2,x_4,\ldots,x_{2k}$, and similarly ${\bf x}_{\text{odd}}=x_1,x_3,\ldots,x_{2k-1}$. }
\begin{alignat}{2}
	&\mathfrak{c}_O:\quad \text{skel}^{\downarrow}({\bf x}_{\text{odd}})&&\text{ constant},\\
	&\mathfrak{c}_E:\quad \text{skel}^{\uparrow}({\bf x}_{\text{even}})&&\text{ constant}.
\end{alignat}
The full chambers are then retrieved as the intersection of two half-chambers. For example, in the language of half-chambers, \eqref{eq:chams_8_exam} becomes  
\begin{align}
	\mathfrak{c}_{T_{ij}}=\mathfrak{c}_{O_{i}} \cap \mathfrak{c}_{E_{j}}, \ \text{ alternatively }\  \mathfrak{c}_{O_i} = \mathfrak{c}_{T_{i1}} \cup \mathfrak{c}_{T_{i2}} \ \text{ and } \  \mathfrak{c}_{E_j} = \mathfrak{c}_{T_{1j}} \cup \mathfrak{c}_{T_{2j}}.
\end{align}
This gives us an alternative \emph{half-chamber} decomposition of the tree-level ABJM amplituhedron form 
\begin{align}
	\Omega[ \mathcal{O}_k] = \sum_{O} \Omega[\mathfrak{c}_O]=\sum_{E} \Omega[\mathfrak{c}_E].
\end{align}
This discussion is easily extended to all loops. When the full $L$-loop chamber was defined, we were concerned with the combinatorial structure of the $(L+1)$-loop fiber \eqref{eq:loop-fiber-def}. More precisely, this takes into account how the lightcone of $y_{L+1} \in \Delta({\bf x})$ intersects the lightcones of any of the $x_i$ or the remaining loop variables $y_a$. To define odd and even half-chambers for higher loops, we instead care only about the intersections of the past and future lightcones of $\mathcal{N}_{y_{L+1}}$. This leads to the notion which we refer to as odd/even \emph{$L$-loop half-chambers}: regions of $\mathcal{O}^{(L)}_k$ for which \begin{alignat}{2}
	&\mathfrak{c}_O^{(L)}:\quad \text{skel}^{\downarrow}({\bf x}_{\text{odd}};y_1,\ldots,y_L)&&\text{ constant},\\
	&\mathfrak{c}_E^{(L)}:\quad \text{skel}^{\uparrow}({\bf x}_{\text{even}};y_1,\ldots,y_L)&&\text{ constant}.
\end{alignat}

\subsection{One-loop half-chambers}
In this section we study the geometry of one-loop half-chambers. For simplicity, we restrict our attention to odd half-chambers, the results immediately generalise to the even half-chambers.

\begin{figure}
	\centering
	\begin{minipage}{0.45\textwidth}
		\centering
		\includegraphics[width=\textwidth]{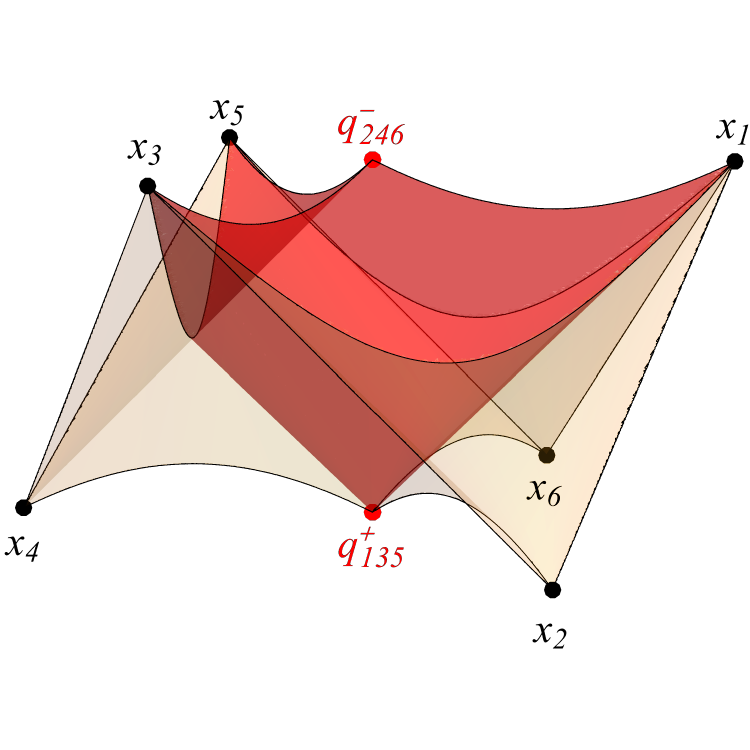}
	\end{minipage}
	\hfill 
	\begin{minipage}{0.45\textwidth}
		\centering
		\includegraphics[width=\textwidth]{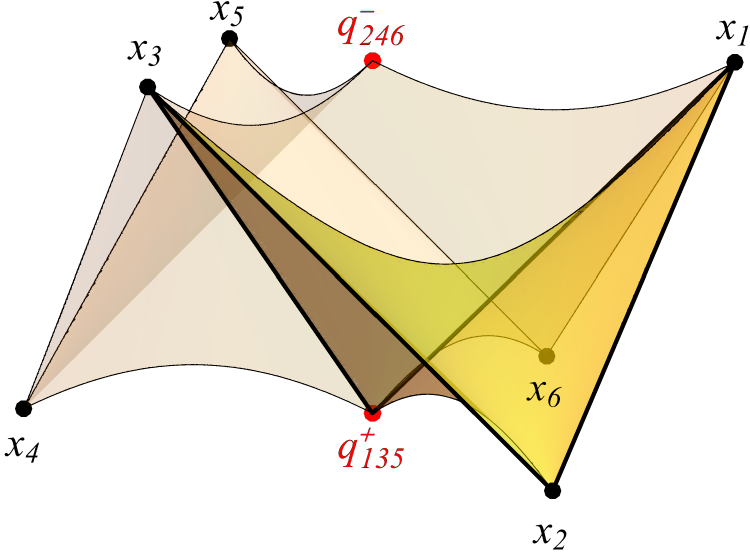}
	\end{minipage}
	\caption{The lightcone of the vertex $q^+_{135}$ divides $\Delta_6(\mathbf{x})$ into four regions. The negative part $\Delta_{135}^-$ is highlighted in red on the left, and part of the positive contribution $\Delta_{\widebar{13}5}=\Delta(x_1,x_2,x_3,q^+_{135})\subset \Delta_{135}^+$ is highlighted in yellow on the right.}
	\label{fig:six_point_q135_lightcone}
\end{figure}


\subsubsection{Six-points}\label{sec:six-point-half-chambers}
Before moving on to a general description of one-loop half-chambers, we will build an intuition by investigating the six-point example in detail. Regardless of the tree-level data chosen, at six-points, the one-loop fiber contains six vertices given by the cusps of the null polygon ${\bf x}$ together with the two triple-cut points $q^+_{135}$ and $q^-_{246}$. Therefore, it suffices to fix the null polygon ${\bf x}$ once and for all and simply vary $y_1 \in \Delta({\bf x})$. To probe the structure of the one-loop half-chambers we must determine how the vertices of the two-loop fiber $\Delta({\bf x};y_1)$ coming from intersections of past lightcones change as we vary $y_1$. 

\begin{figure}
	\centering
	\includegraphics[width=0.16\textwidth]{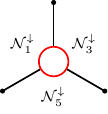} \quad \quad \quad
	\includegraphics[width=0.16\textwidth]{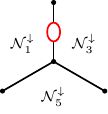} \quad \quad \quad
	\includegraphics[width=0.16\textwidth]{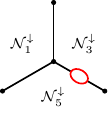} \quad \quad \quad
	\includegraphics[width=0.16\textwidth]{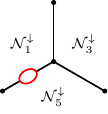}  
	\caption{A local view of the vertex $q^+_{135} \in \Delta({\bf x})$ at six-points where the lightcone of the point $y_1 \in \Delta({\bf x})$ is depicted in red. Depending on the position of $y_1$ its null-cone will intersect in either of the four configurations depicted above. The first diagram corresponds to $(y_1-q^+_{135})^2<0$ whilst the remaining three terms correspond to $(y_1-q^+_{135})^2>0$.}
	\label{fig:vert_configs}
\end{figure}

To begin, we focus on a single point $q^+_{135}$ and determine whether it is a vertex of the geometry $\Delta({\bf x};y_1)$. Suppose $(y_1-q^+_{135})^2<0$, then since $\Delta({\bf x};y_1)$ is defined to be all points $y_2 \in \Delta({\bf x})$ positively separated from $y_1$, this implies that $q^+_{135} $ is not a vertex of the two-loop fiber. In this case the lightcone of $y_1$ \emph{cuts off} the vertex $q^+_{135}$ to generate three new intersection points $q^+_{13y_1}$, $q^+_{35y_1}$ and $q^+_{53y_1}$ as depicted in the first diagram of Fig. \ref{fig:vert_configs}. Instead, if we now take $y_1$ such that $(y_1-q^+_{135})^2>0$, we will find that $q^+_{135}$ is a vertex of the two-loop fiber. However, there are three possible configurations in which this can happen, also depicted in Fig. \ref{fig:vert_configs}, where the lightcone of $y_1$ cuts one of the edges $\mathcal{N}_1 \cap \mathcal{N}_3$, $\mathcal{N}_3 \cap \mathcal{N}_5$ or $\mathcal{N}_5 \cap \mathcal{N}_1$ generating two new intersection points $q^\pm_{13y_1}$, $q^\pm_{35y_1}$ or $q^\pm_{51y_1}$, respectively.  

These four cases can be realised geometrically by \emph{refining} the one-loop fiber $\Delta({\bf x})$ by the lightcone of $q^+_{135}$. This naturally decomposes $\Delta({\bf x})$ into a region $\Delta^+_{135}$ which is positively separated from $q^+_{135}$, and a region $\Delta^-_{135}$ which is negatively separated from $q^+_{135}$:
\begin{align}
	\Delta({\bf x}) = \Delta^+_{135} \cup \Delta^-_{135}.
\end{align}
The positive part $\Delta_{135}^+$ further splits into three regions which are geometrically four-point one-loop fibers. Explicitly, we have 
\begin{align}
	\Delta_{135}^+=\Delta_{\widebar{13}5}\cup \Delta_{1\widebar{35}}\cup\Delta_{\widebar{1}3\widebar{5}}=\Delta(x_1,x_2,x_3,q^+_{135})\cup\Delta(x_3,x_4,x_5,q^+_{135})\cup\Delta(x_5,x_6,x_1,q^+_{135})\,.
\end{align}
This decomposition of $\Delta({\bf x})$ is depicted in Fig. \ref{fig:six_point_q135_lightcone}. 
If the loop variable $y_1$ is inside one of the four regions $\Delta_{135}^-,\; \Delta_{\widebar{1}3\widebar{5}},\; \Delta_{\widebar{13}5},\; \Delta_{1\widebar{35}}$, then the two-loop fiber $\Delta({\mathbf{x};y_1})$ will precisely have the structure depicted in Fig. \ref{fig:vert_configs}, respectively.

Thus, at six points there are four one-loop odd half-chambers, and the skeletal structure of the associated two-loop fiber takes the form of a planar three-point one-loop Feynman diagram (without tadpoles), as depicted in Fig. \ref{fig:vert_configs}. This is reminiscent of the skeletal structure of $n$-point one-loop fiber, which takes the form of planar tree-level $n/2$-points cubic scalar Feynman diagrams. 

At tree-level, we further found it useful to consider the dual of these diagrams, which are triangulations of $n/2$-gons. The analogous structure for these six-point one-loop half-chambers are triangulations of a triangle (or rather `scaffolding' triangulations of a hexagon) with one internal puncture, which we denote graphically as
\begin{align}
	\mathcal{O}_3^{(1)}= \ &\raisebox{-28pt}{\includegraphics[scale=0.8]{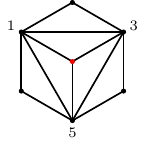}}\cup\raisebox{-28pt}{\includegraphics[scale=0.8]{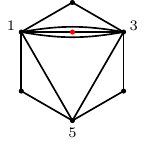}}\cup\raisebox{-28pt}{\includegraphics[scale=0.8]{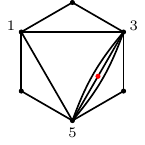}}\cup\raisebox{-28pt}{\includegraphics[scale=0.8]{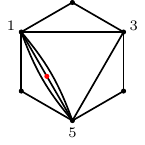}}\,,
\end{align}
where the triangulations in this union are dual to the diagrams in Fig. \ref{fig:vert_configs}. 

The canonical form of the lightcone geometries $\Delta_{135}^-,\; \Delta_{\widebar{1}3\widebar{5}},\; \Delta_{\widebar{13}5},\; \Delta_{1\widebar{35}}$ can be found by summing over their vertices in precisely the same way as we discussed for the one-loop fiber in section \ref{sec:tree-level-chambers}. Explicitly,
\begin{align}
	\Omega[\Delta_{\widebar{13}5}]&=\omega_{123}-\omega_{23q^+_{135}}+\omega_{3q^+_{135}1}-\omega_{q^+_{135}12},\\
	\Omega[\Delta_{1\widebar{35}}]&=\omega_{345}-\omega_{34q^+_{135}}+\omega_{4q^+_{135}5}-\omega_{q^+_{135}51},\\
	\Omega[\Delta_{\widebar{1}3\widebar{5}}]&=\omega_{512}-\omega_{51q^+_{135}}+\omega_{1q^+_{135}2}-\omega_{q^+_{135}25},\\
	\Omega[\Delta_{135}^-]&=\omega^-_{246}-\omega^-_{24q^+_{135}}+\omega^-_{4q^+_{135}6}-\omega^-_{q^+_{135}62}\,.
\end{align}
When adding these together we recover the canonical form of the full one-loop fiber, as expected:
\begin{align}
	\Omega[\Delta({\bf x})]=\Omega[\Delta_{\widebar{13}5}]+\Omega[\Delta_{1\widebar{35}}]+\Omega[\Delta_{\widebar{1}3\widebar{5}}]+	\Omega[\Delta_{135}^-].
\end{align}

\subsubsection{Eight points}\label{sec:8pt-half-chamber}
\begin{figure}[h]
	\centering
	\includegraphics[width=0.9\textwidth]{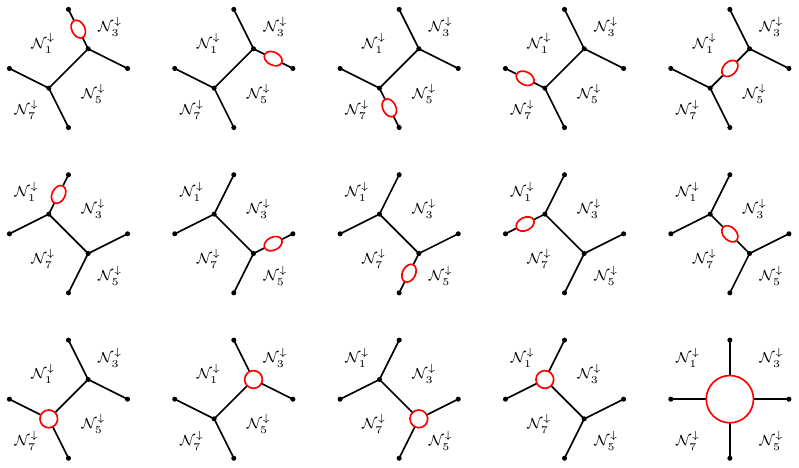} 
	\caption{The $15$ possible past lightcone configurations for the two-loop fiber. }
	\label{fig:8point-half-chambers} 
\end{figure}

At eight-points there are two tree-level half-chambers, which means that we can no longer discover all two-loop fibers by fixing ${\bf x}$ and varying $y_1$. Instead, we need to study the structure of $\Delta({\bf x};y_1)$ by probing $y_1$ over $\Delta({\bf x})$ for ${\bf x}$ in both of the tree-level half-chambers. An explicit check yields 15 distinct one-loop half-chambers. The combinatorial structures $\text{skel}^{\downarrow}({\bf x}_{\text{odd}};y_1)$ are recorded in Fig. \ref{fig:8point-half-chambers}. We again find that the past vertices of the two-loop fiber take the form of a one-loop Feynman diagram for four-point $\Tr(\phi^3)$ amplitudes.

If we had instead fixed ${\bf x}$ in either one of the two half-chambers and then studied the structure of the two-loop fiber as we varied $y_1$, then we would have identified only 8 of the 15 half-chambers. To be more precise, if we fix ${\bf x}$ in the half-chamber whose one-loop fiber contains $q^+_{137},q^+_{357}$, then we will never be able to find a two-loop fiber which has the vertex $q^+_{135}$ or $q^+_{157}$. Instead, we find only two-loop fibers $\Delta({\bf x};y_1)$ which are \emph{compatible} with the one-loop fiber $\Delta({\bf x})$.

Using our graphical notation, the odd half-chamber decomposition of the one-loop amplituhedron takes the following form
\begin{align}\label{eq:8pt-half-chamber-sum}
\mathcal{O}_{4}^{(1)} = \ &\vcenter{\hbox{\includegraphics[scale=0.75]{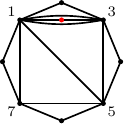}}}\,\cup\,\vcenter{\hbox{\includegraphics[scale=0.75]{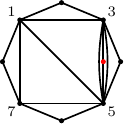}}}\,\cup\,\vcenter{\hbox{\includegraphics[scale=0.75]{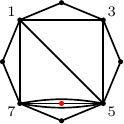}}}\,\cup\,\vcenter{\hbox{\includegraphics[scale=0.75]{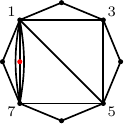}}}\,\cup \vcenter{\hbox{\includegraphics[scale=0.75]{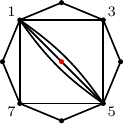}}}\notag \\
&\vcenter{\hbox{\includegraphics[scale=0.75]{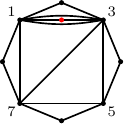}}}\,\cup\,\vcenter{\hbox{\includegraphics[scale=0.75]{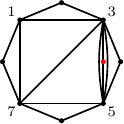}}}\,\cup\,\vcenter{\hbox{\includegraphics[scale=0.75]{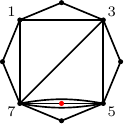}}}\,\cup\,\vcenter{\hbox{\includegraphics[scale=0.75]{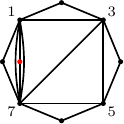}}}\,\cup \vcenter{\hbox{\includegraphics[scale=0.75]{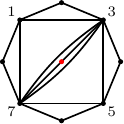}}}\notag \\
& \vcenter{\hbox{\includegraphics[scale=0.75]{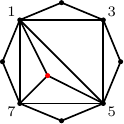}}}\, \cup \vcenter{\hbox{\includegraphics[scale=0.75]{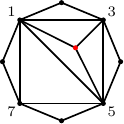}}}\, \cup\,\vcenter{\hbox{\includegraphics[scale=0.75]{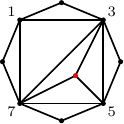}}}\,  \cup  \vcenter{\hbox{\includegraphics[scale=0.75]{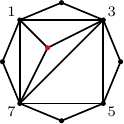}}}\, \cup \,\vcenter{\hbox{\includegraphics[scale=0.75]{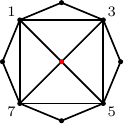}}}.
\end{align}
The above diagrams are dual to the Feynman diagrams (past one-skeletons) appearing in Fig. \ref{fig:8point-half-chambers} and are organised accordingly. 

\subsubsection{General number of particles}\label{sec:generaln}
The technique to find one-loop half-chambers for a general number of particles is similar in spirit to the arguments used in the six-point example. By fixing ${\bf x}$, we find the various associated two-loop fibers by refining $\Delta({\bf x})$ by the lightcones of all the $q^+_{ijk}\in\mathcal{V}[\Delta({\bf x})]$. This creates some number of regions which give rise to different combinatorial structures of $\Delta({\bf x};y_1)$.

However, as we emphasised when studying the eight-point example, we will only discover two-loop fibers which are compatible with the chosen $\Delta({\bf x})$ if we proceed in this manner. Since there are multiple tree-level chambers for $n\geq8$ giving rise to different structures of $\Delta({\bf x})$, we need to sample ${\bf x}$ in all of the (Catalan number many) tree-level half-chambers. This will allow us to find all possible past structures of the two-loop fibers, which defines the full set of one-loop half-chambers.

Similar to the observation around Fig. \ref{fig:six_point_q135_lightcone} for six points, the lightcone of $q^+_{ijk}$ generically refines $\Delta({\bf x})$ into four regions:
\begin{align}
	\Delta({\bf x}) = \Delta_{ijk}^+ \cup \Delta_{ijk}^- = (\Delta_{\widebar{ij}k} \cup \Delta_{\widebar{i}j\widebar{k}} \cup \Delta_{i\widebar{jk}}) \cup \Delta_{ijk}^-,
	\label{eq:decompose_4}
\end{align}
where the three-terms appearing in the positive part $\Delta_{ijk}^+$ correspond to one-loop fibers of lower-point configurations
\begin{align}
	\Delta_{\widebar{ij}k} &=\Delta(x_i,x_{i+1},\ldots,x_j,q^\pm_{ijk}), \notag \\
	\Delta_{i\widebar{jk}} &= \Delta(x_j,x_{j+1},\ldots,x_k,q^\pm_{ijk}), \notag \\
	\Delta_{\widebar{i}j\widebar{k}} &= \Delta(x_k,x_{k+1},\ldots,x_i,q^\pm_{ijk}).
\end{align} 
These regions determine how the past lightcone of $y_1$ intersects the edges meeting at the vertex $q^+_{ijk}$. To determine the vertices of the two-loop fiber when there are multiple vertices $q^+_{ijk} \in \mathcal{V}[\Delta({\bf x})]$, we need to specify in which of these four regions $y_1$ is. Thus, for fixed ${\bf x}$ we can find all possible structures of the two-loop fiber $\Delta({\bf x};y_1)$ by considering the maximal intersections of the regions \eqref{eq:decompose_4} for all vertices $q^+_{ijk} \in \mathcal{V}[\Delta({\bf x})]$.

We start with a point ${\bf x}$ fixed inside a tree-level half-chamber specified by an odd scaffolding triangulation $O$ of an $n$-gon. To refine these by one-loop half-chambers, we consider how each of the $(k-2)$ triple cuts $q^+_{ijk} \in O$ divide $\Delta({\bf x})$ into four regions as in \eqref{eq:decompose_4}, and then take their common refinement. Since each half-chamber is defined as the intersection of $(k-2)$ regions taking the general form
\begin{align}
	\Delta_{\widebar{a_1b_1}c_1} \cap \ldots \cap \Delta_{\widebar{a_mb_m}c_m} \cap \Delta^-_{a_{m+1}b_{m+1}c_{m+1}} \cap \ldots \cap \Delta^-_{a_{k-2}b_{k-2}c_{k-2}},
	\label{eq:gen_odd_hc}
\end{align}
we would naively expect to find $4^{k-2}$ different two-loop fibers. However, many of these intersections will be empty. Furthermore, many of the labels will be redundant as some of the regions strictly contain others. We now move on to describe the set of non-empty intersections in detail.

We first address the problem of redundancy of labels. In order to do so, we must determine when a region in \eqref{eq:gen_odd_hc} is a subset of another. This only occurs when we consider the positive parts with respect to a given triple-cut point. It is straightforward to see that one region $\Delta_{\widebar{ij}k}$ is a subset of another $\Delta_{\widebar{lm}n}$ when we have that
\begin{align}
	\{i,i+1,\ldots,j\} \subset \{l,l+1,\ldots,m \}.
\end{align}
We now turn to characterising the set of non-empty intersections. First we inscribe an $m$-gon with vertex labels $\{ b_1,b_2,\ldots,b_m\}$ inside the tree-level triangulation $O$ using pre-existing edges. This will split the set of triangles appearing in $O$ into two subsets, those inside the $m$-gon, which we shall denote $Q$, and those outside the $m$-gon, which we shall denote $\bar{Q}$. From the triangles contained in $\bar{Q}$ there will be $m$ that share a boundary with $Q$ which take the form $t_{b_{i}e_ib_{i+1}} \in \bar{Q}$. All non-empty intersections then simply take the form
\begin{align}\label{eq:region-to-half-chamber}
	\left(\bigcap_{t_{ijk} \in Q} \Delta_{ijk}^-\right) \cap \left(\bigcap_{i=1}^m  \Delta_{\widebar{b_i}e_{i}\widebar{b_{i+1}}}\right)\to 	\vcenter{\hbox{\includegraphics[width=35mm]{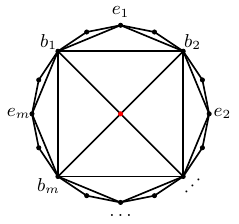}}}.
\end{align}
Here we have introduced our familiar graphical notation for the one-loop half-chambers. The region on the left of this equation gives rise to a two-loop fiber whose odd structure is dual to the diagram on the right. We arrive at this diagram by starting from the odd tree-level triangulation, removing all edges contained in the sub-polygon $Q$ and introducing a red vertex connected to all vertices of $Q$. The remainder of the triangulation is left unchanged. In this way, each one-loop chamber is labelled by a triangulation of a once punctured $k$-gon. If it is compatible with any other tree-level triangulation, \textit{i.e.} only triangles appearing in $O$ will appear in the triangulation for the one-loop odd half-chamber, then it will also refine this one-loop fiber. Note, this procedure never produces a triangulation whose dual Feynman diagram contains a tadpole. Thus, $n$-point one-loop half-chambers are labelled by triangulations of an $n/2$-gon with one internal puncture, and they will refine a given one-loop fiber exactly when it is compatible with the underlying tree-level triangulation.

For each given one-loop fiber $\Delta({\bf x})$, equation \eqref{eq:region-to-half-chamber} defines a region inside $\Delta({\bf x})$ which gives rise to a compatible two-loop fiber. By definition, the structure of these one-loop fibers remains constant over a tree-level chamber, however the same is not true for these regions. A further refinement of the tree-level chambers is needed to determine their structure. We will explain how this refinement works in appendix \ref{sec:app_refinement}.

\paragraph{Examples.}
Let us return to the eight-point example of section \ref{sec:8pt-half-chamber}. We start from a point ${\bf x}_*$ inside the tree-level half-chamber $O=\{t_{137},t_{357}\}$ with one-loop fiber $\Delta({\bf x}_*)$. The one-loop fiber is refined by the lightcones $\mathcal{N}_{q^+_{137}}$ and $\mathcal{N}_{q^+_{357}}$ as
\begin{align}
	\Delta({\bf x}_*) &= \Delta_{\widebar{13}7} \cup \Delta_{\widebar{35}7} \cup \Delta_{3\widebar{57}} \cup \Delta_{\widebar{1}3\widebar{7}} \cup \left( \Delta_{1\widebar{37}} \cap \Delta_{\widebar{3}5\widebar{7}} \right) \notag \\
	&\cup  \left( \Delta_{137} \cap\Delta_{\widebar{3}5\widebar{7}}\right) \cup \left(  \Delta_{357} \cap \Delta_{1\widebar{37}}\right) \cup  \left( \Delta_{137} \cap \Delta_{357} \right)\,.
\end{align}
Using our graphical notation, the above refinement of the one-loop fiber reads
\begin{align}\label{gen:eq1}
\Delta({\bf x}_*) = \ &\vcenter{\hbox{\includegraphics[scale=0.75]{Figures/eight_point_hc_1}}}\,\cup\,\vcenter{\hbox{\includegraphics[scale=0.75]{Figures/eight_point_hc_2}}}\,\cup\,\vcenter{\hbox{\includegraphics[scale=0.75]{Figures/eight_point_hc_3}}}\,\cup\,\vcenter{\hbox{\includegraphics[scale=0.75]{Figures/eight_point_hc_4}}}\,\cup \vcenter{\hbox{\includegraphics[scale=0.75]{Figures/eight_point_hc_5}}}\notag \\
& \vcenter{\hbox{\includegraphics[scale=0.75]{Figures/eight_point_hc_6}}}\, \cup\,\vcenter{\hbox{\includegraphics[scale=0.75]{Figures/eight_point_hc_7}}}\, \cup \,\vcenter{\hbox{\includegraphics[scale=0.75]{Figures/eight_point_hc_8}}}.
\end{align}
Had we began with a point ${\bf x}_*'$ in the tree-level chamber labelled by $O=\{t_{135},t_{157}\}$ we would have instead found
\begin{align}\label{gen:eq2}
\Delta({\bf x}_*') =  \ &\vcenter{\hbox{\includegraphics[scale=0.75]{Figures/eight_point_hc_new_3}}}\,\cup\,\vcenter{\hbox{\includegraphics[scale=0.75]{Figures/eight_point_hc_new_4}}}\,\cup\,\vcenter{\hbox{\includegraphics[scale=0.75]{Figures/eight_point_hc_new_5}}}\,\cup\,\vcenter{\hbox{\includegraphics[scale=0.75]{Figures/eight_point_hc_new_6}}}\,\cup \vcenter{\hbox{\includegraphics[scale=0.75]{Figures/eight_point_hc_new_7}}}\notag \\
&\vcenter{\hbox{\includegraphics[scale=0.75]{Figures/eight_point_hc_new_1}}}\, \cup\,\vcenter{\hbox{\includegraphics[scale=0.75]{Figures/eight_point_hc_new_2}}}\, \cup \,\vcenter{\hbox{\includegraphics[scale=0.75]{Figures/eight_point_hc_8}}}.
\end{align}
Importantly, the labels in \eqref{gen:eq1} and \eqref{gen:eq2} should be interpreted as regions of the one-loop fiber $\Delta({\bf x})$, with ${\bf x}$ held fixed, such that the structure of the two-loop fiber reproduces the corresponding diagram. Consequently, the associated canonical forms depend solely on the loop variable $y_1$. In contrast, the diagrams in \eqref{eq:8pt-half-chamber-sum} represent regions of the full one-loop amplituhedron $\mathcal{O}^{(1)}_k$, and thus their canonical forms depend on both the tree-level ${\bf x}$ and the loop-level $y_1$ variables.

\subsection{Higher loop half-chambers}

Much of the structure of one-loop half-chambers can be generalised to higher loops as well. The most striking general structure comes from the way we label the half-chambers. We have argued that tree-level half-chambers can be labelled by cubic tree Feynman diagrams with $n/2$ particles, and similarly one-loop half-chambers are labelled by cubic one-loop Feynman diagrams. In general, the $L$-loop half-chambers are captured by (tadpole-free) $L$-loop cubic Feynman diagrams on $n/2$ particles. Furthermore, these labels directly describe half of the combinatorial structure of the $(L+1)$-loop fiber. We sometimes prefer to think in dual diagrams, in which case the half-chambers are instead labelled by odd/even triangulations of $n$-gons with $L$ punctures.

The fact that the half-chambers for ABJM are labelled by Feynman diagrams for a cubic scalar theory with half the number of particles is at first sight very surprising. However, we can gain some intuition for this statement from the geometry. From the definition \eqref{eq:loop-fiber-def}, it is clear that we can obtain every $(L+1)$-loop fiber $\Delta({\bf x};y_1,\ldots,y_L)$ from the $L$-loop fiber $\Delta({\bf x};y_1,\ldots,y_{L-1})$ and intersecting with $\mathcal{N}_{y_L}^+$. The argument now follows an inductive reasoning where we assume that the past boundaries of $L$-loop fiber indeed look like an $(L-1)$-loop Feynman diagram. Intersecting this with $\mathcal{N}_{y_L}^+$ will cut away a region contained in the past lightcone of $y_L$, which will induce a new loop in the diagram.

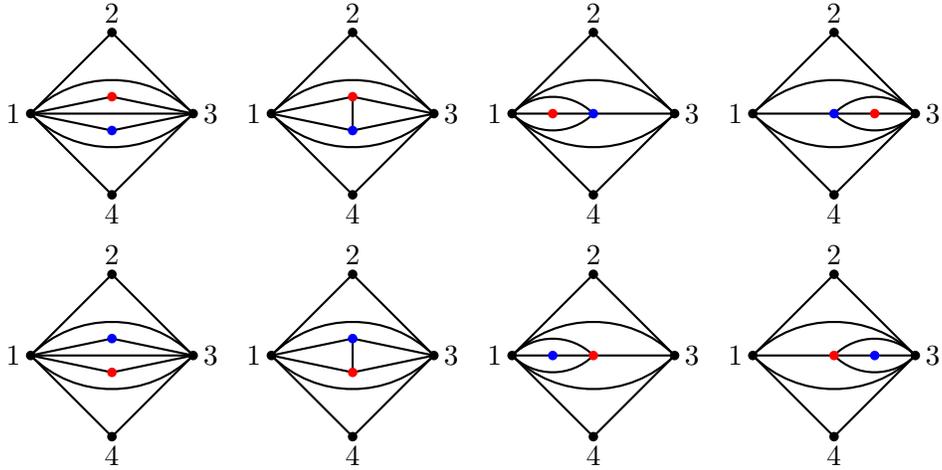
\begin{figure}
	\centering
	\begin{tikzpicture}[scale=0.9]

		\newcommand{\chamfourtadpole}[3]{
			\pgfmathsetmacro{\cx}{#1}  
			\pgfmathsetmacro{\cy}{#2} 
			\pgfmathsetmacro{\r}{#3}   
			
			\pgfmathsetmacro{\xA}{\cx-\r}
			\pgfmathsetmacro{\yA}{\cy}
			\pgfmathsetmacro{\xB}{\cx + \r}
			\pgfmathsetmacro{\yB}{\cy}
			\pgfmathsetmacro{\nx}{\cx}
			\pgfmathsetmacro{\ny}{\cy+\r}
			\pgfmathsetmacro{\nx}{\cx}
			\pgfmathsetmacro{\ny}{\cy+\r}
			\pgfmathsetmacro{\sx}{\cx}
			\pgfmathsetmacro{\sy}{\cy-\r}

			\node at (\xA,\yA) [left] {1};
			\node at (\xB,\yB) [right] {3};
			\draw[thick] (\xA,\yA) -- (\cx,\cy) -- (\xB,\yB);
			\draw[thick] (\xA,\yA) -- (\nx,\ny) -- (\xB,\yB) --(\sx,\sy) -- (\xA,\yA);
			\draw[thick] (\xA,\yA) -- (0,-0.25) -- (\xB,\yB);
			\draw[thick] (\xA,\yA) -- (0,0.25) -- (\xB,\yB);
			\fill (\xA,\yA) circle (2pt);
			\fill (\xB,\yB) circle (2pt);
			\fill (\nx,\ny) circle (2pt);
			\fill (\sx,\sy) circle (2pt);
			\draw[thick, black] (\xA,\yA) to[out=45, in=90+45] (\xB,\yB);
			\draw[thick, black] (\xA,\yA) to[out=270+45, in=270-45] (\xB,\yB);
			\node at (0,\r) [above] {2};
			\node at (0,-\r) [below] {4};
			\fill[blue] (0,-0.25) circle (2pt);
			\fill[red] (0,0.25) circle (2pt);
		}
		\chamfourtadpole{0}{0}{1.2}
	\end{tikzpicture}
	\begin{tikzpicture}[scale=0.9]

		\newcommand{\chamfourtadpole}[3]{
			\pgfmathsetmacro{\cx}{#1}  
			\pgfmathsetmacro{\cy}{#2}  
			\pgfmathsetmacro{\r}{#3}   
			
			\pgfmathsetmacro{\xA}{\cx-\r}
			\pgfmathsetmacro{\yA}{\cy}
			\pgfmathsetmacro{\xB}{\cx + \r}
			\pgfmathsetmacro{\yB}{\cy}
			\pgfmathsetmacro{\nx}{\cx}
			\pgfmathsetmacro{\ny}{\cy+\r}
			\pgfmathsetmacro{\nx}{\cx}
			\pgfmathsetmacro{\ny}{\cy+\r}
			\pgfmathsetmacro{\sx}{\cx}
			\pgfmathsetmacro{\sy}{\cy-\r}

			\node at (\xA,\yA) [left] {1};
			\node at (\xB,\yB) [right] {3};
			\draw[thick] (\xA,\yA) -- (\nx,\ny) -- (\xB,\yB) --(\sx,\sy) -- (\xA,\yA);
			\draw[thick] (\xA,\yA) -- (0,-0.25) -- (\xB,\yB);
			\draw[thick] (\xA,\yA) -- (0,0.25) -- (\xB,\yB);
			\draw[thick] (0,-0.25) -- (0,0.25);
			\fill (\xA,\yA) circle (2pt);
			\fill (\xB,\yB) circle (2pt);
			\fill (\nx,\ny) circle (2pt);
			\fill (\sx,\sy) circle (2pt);
			\draw[thick, black] (\xA,\yA) to[out=45, in=90+45] (\xB,\yB);
			\draw[thick, black] (\xA,\yA) to[out=270+45, in=270-45] (\xB,\yB);
			\node at (0,\r) [above] {2};
			\node at (0,-\r) [below] {4};
			\fill[blue] (0,-0.25) circle (2pt);
			\fill[red] (0,0.25) circle (2pt);
		}
		\chamfourtadpole{0}{0}{1.2}
	\end{tikzpicture}
	\begin{tikzpicture}[scale=0.9]

		\newcommand{\chamfourtadpole}[3]{
			\pgfmathsetmacro{\cx}{#1}  %
			\pgfmathsetmacro{\cy}{#2}  %
			\pgfmathsetmacro{\r}{#3}   %

			\pgfmathsetmacro{\xA}{\cx-\r}
			\pgfmathsetmacro{\yA}{\cy}
			\pgfmathsetmacro{\xB}{\cx + \r}
			\pgfmathsetmacro{\yB}{\cy}
			\pgfmathsetmacro{\nx}{\cx}
			\pgfmathsetmacro{\ny}{\cy+\r}
			\pgfmathsetmacro{\nx}{\cx}
			\pgfmathsetmacro{\ny}{\cy+\r}
			\pgfmathsetmacro{\sx}{\cx}
			\pgfmathsetmacro{\sy}{\cy-\r}

			\node at (\xA,\yA) [left] {1};
			\node at (\xB,\yB) [right] {3};
			\draw[thick] (\xA,\yA) -- (\cx,\cy) -- (\xB,\yB);
			\draw[thick] (\xA,\yA) -- (\nx,\ny) -- (\xB,\yB) --(\sx,\sy) -- (\xA,\yA);
			\fill (\xA,\yA) circle (2pt);
			\fill (\xB,\yB) circle (2pt);
			\fill (\nx,\ny) circle (2pt);
			\fill (\sx,\sy) circle (2pt);
			\draw[thick, black] (\xA,\yA) to[out=45, in=90+45] (\xB,\yB);
			\draw[thick, black] (\xA,\yA) to[out=270+45, in=270-45] (\xB,\yB);
			\draw[thick, black] (\xA,\yA) to[out=45, in=90+45] (\cx,\cy);
			\draw[thick, black] (\xA,\yA) to[out=270+45, in=270-45] (\cx,\cy);
			\node at (0,\r) [above] {2};
			\node at (0,-\r) [below] {4};
			\fill[blue] (\cx,\cy) circle (2pt);
			\fill[red] ($0.5*(\xA,\yA)$) circle (2pt);
		}
		\chamfourtadpole{0}{0}{1.2}
	\end{tikzpicture}
	\begin{tikzpicture}[scale=0.9]

		\newcommand{\chamfourtadpole}[3]{
			\pgfmathsetmacro{\cx}{#1}  %
			\pgfmathsetmacro{\cy}{#2}  %
			\pgfmathsetmacro{\r}{#3}   %

			\pgfmathsetmacro{\xA}{\cx-\r}
			\pgfmathsetmacro{\yA}{\cy}
			\pgfmathsetmacro{\xB}{\cx + \r}
			\pgfmathsetmacro{\yB}{\cy}
			\pgfmathsetmacro{\nx}{\cx}
			\pgfmathsetmacro{\ny}{\cy+\r}
			\pgfmathsetmacro{\nx}{\cx}
			\pgfmathsetmacro{\ny}{\cy+\r}
			\pgfmathsetmacro{\sx}{\cx}
			\pgfmathsetmacro{\sy}{\cy-\r}

			\node at (\xA,\yA) [left] {1};
			\node at (\xB,\yB) [right] {3};
			\draw[thick] (\xA,\yA) -- (\cx,\cy) -- (\xB,\yB);
			\draw[thick] (\xA,\yA) -- (\nx,\ny) -- (\xB,\yB) --(\sx,\sy) -- (\xA,\yA);
			\fill (\xA,\yA) circle (2pt);
			\fill (\xB,\yB) circle (2pt);
			\fill (\nx,\ny) circle (2pt);
			\fill (\sx,\sy) circle (2pt);
			\draw[thick, black] (\xA,\yA) to[out=45, in=90+45] (\xB,\yB);
			\draw[thick, black] (\xA,\yA) to[out=270+45, in=270-45] (\xB,\yB);
			\draw[thick, black] (\cx,\cy) to[out=45, in=90+45] (\xB,\yB);
			\draw[thick, black] (\cx,\cy) to[out=270+45, in=270-45] (\xB,\yB);
			\node at (0,\r) [above] {2};
			\node at (0,-\r) [below] {4};
			\fill[blue] (\cx,\cy) circle (2pt);
			\fill[red] ($0.5*(\xB,\yB)$) circle (2pt);
		}
		\chamfourtadpole{0}{0}{1.2}
	\end{tikzpicture}\\
	\begin{tikzpicture}[scale=0.9]

		\newcommand{\chamfourtadpole}[3]{
			\pgfmathsetmacro{\cx}{#1}  %
			\pgfmathsetmacro{\cy}{#2}  %
			\pgfmathsetmacro{\r}{#3}   %

			\pgfmathsetmacro{\xA}{\cx-\r}
			\pgfmathsetmacro{\yA}{\cy}
			\pgfmathsetmacro{\xB}{\cx + \r}
			\pgfmathsetmacro{\yB}{\cy}
			\pgfmathsetmacro{\nx}{\cx}
			\pgfmathsetmacro{\ny}{\cy+\r}
			\pgfmathsetmacro{\nx}{\cx}
			\pgfmathsetmacro{\ny}{\cy+\r}
			\pgfmathsetmacro{\sx}{\cx}
			\pgfmathsetmacro{\sy}{\cy-\r}

			\node at (\xA,\yA) [left] {1};
			\node at (\xB,\yB) [right] {3};
			\draw[thick] (\xA,\yA) -- (\cx,\cy) -- (\xB,\yB);
			\draw[thick] (\xA,\yA) -- (\nx,\ny) -- (\xB,\yB) --(\sx,\sy) -- (\xA,\yA);
			\draw[thick] (\xA,\yA) -- (0,-0.25) -- (\xB,\yB);
			\draw[thick] (\xA,\yA) -- (0,0.25) -- (\xB,\yB);
			\fill (\xA,\yA) circle (2pt);
			\fill (\xB,\yB) circle (2pt);
			\fill (\nx,\ny) circle (2pt);
			\fill (\sx,\sy) circle (2pt);
			\draw[thick, black] (\xA,\yA) to[out=45, in=90+45] (\xB,\yB);
			\draw[thick, black] (\xA,\yA) to[out=270+45, in=270-45] (\xB,\yB);
			\node at (0,\r) [above] {2};
			\node at (0,-\r) [below] {4};
			\fill[red] (0,-0.25) circle (2pt);
			\fill[blue] (0,0.25) circle (2pt);
		}
		\chamfourtadpole{0}{0}{1.2}
	\end{tikzpicture}
	\begin{tikzpicture}[scale=0.9]

		\newcommand{\chamfourtadpole}[3]{
			\pgfmathsetmacro{\cx}{#1}  %
			\pgfmathsetmacro{\cy}{#2}  %
			\pgfmathsetmacro{\r}{#3}   %

			\pgfmathsetmacro{\xA}{\cx-\r}
			\pgfmathsetmacro{\yA}{\cy}
			\pgfmathsetmacro{\xB}{\cx + \r}
			\pgfmathsetmacro{\yB}{\cy}
			\pgfmathsetmacro{\nx}{\cx}
			\pgfmathsetmacro{\ny}{\cy+\r}
			\pgfmathsetmacro{\nx}{\cx}
			\pgfmathsetmacro{\ny}{\cy+\r}
			\pgfmathsetmacro{\sx}{\cx}
			\pgfmathsetmacro{\sy}{\cy-\r}

			\node at (\xA,\yA) [left] {1};
			\node at (\xB,\yB) [right] {3};
			\draw[thick] (\xA,\yA) -- (\nx,\ny) -- (\xB,\yB) --(\sx,\sy) -- (\xA,\yA);
			\draw[thick] (\xA,\yA) -- (0,-0.25) -- (\xB,\yB);
			\draw[thick] (\xA,\yA) -- (0,0.25) -- (\xB,\yB);
			\draw[thick] (0,-0.25) -- (0,0.25);
			\fill (\xA,\yA) circle (2pt);
			\fill (\xB,\yB) circle (2pt);
			\fill (\nx,\ny) circle (2pt);
			\fill (\sx,\sy) circle (2pt);
			\draw[thick, black] (\xA,\yA) to[out=45, in=90+45] (\xB,\yB);
			\draw[thick, black] (\xA,\yA) to[out=270+45, in=270-45] (\xB,\yB);
			\node at (0,\r) [above] {2};
			\node at (0,-\r) [below] {4};
			\fill[red] (0,-0.25) circle (2pt);
			\fill[blue] (0,0.25) circle (2pt);
		}
		\chamfourtadpole{0}{0}{1.2}
	\end{tikzpicture}
	\begin{tikzpicture}[scale=0.9]

		\newcommand{\chamfourtadpole}[3]{
			\pgfmathsetmacro{\cx}{#1}  %
			\pgfmathsetmacro{\cy}{#2}  %
			\pgfmathsetmacro{\r}{#3}   %

			\pgfmathsetmacro{\xA}{\cx-\r}
			\pgfmathsetmacro{\yA}{\cy}
			\pgfmathsetmacro{\xB}{\cx + \r}
			\pgfmathsetmacro{\yB}{\cy}
			\pgfmathsetmacro{\nx}{\cx}
			\pgfmathsetmacro{\ny}{\cy+\r}
			\pgfmathsetmacro{\nx}{\cx}
			\pgfmathsetmacro{\ny}{\cy+\r}
			\pgfmathsetmacro{\sx}{\cx}
			\pgfmathsetmacro{\sy}{\cy-\r}

			\node at (\xA,\yA) [left] {1};
			\node at (\xB,\yB) [right] {3};
			\draw[thick] (\xA,\yA) -- (\cx,\cy) -- (\xB,\yB);
			\draw[thick] (\xA,\yA) -- (\nx,\ny) -- (\xB,\yB) --(\sx,\sy) -- (\xA,\yA);
			\fill (\xA,\yA) circle (2pt);
			\fill (\xB,\yB) circle (2pt);
			\fill (\nx,\ny) circle (2pt);
			\fill (\sx,\sy) circle (2pt);
			\draw[thick, black] (\xA,\yA) to[out=45, in=90+45] (\xB,\yB);
			\draw[thick, black] (\xA,\yA) to[out=270+45, in=270-45] (\xB,\yB);
			\draw[thick, black] (\xA,\yA) to[out=45, in=90+45] (\cx,\cy);
			\draw[thick, black] (\xA,\yA) to[out=270+45, in=270-45] (\cx,\cy);
			\node at (0,\r) [above] {2};
			\node at (0,-\r) [below] {4};
			\fill[red] (\cx,\cy) circle (2pt);
			\fill[blue] ($0.5*(\xA,\yA)$) circle (2pt);
		}
		\chamfourtadpole{0}{0}{1.2}
	\end{tikzpicture}
	\begin{tikzpicture}[scale=0.9]

		\newcommand{\chamfourtadpole}[3]{
			\pgfmathsetmacro{\cx}{#1}  %
			\pgfmathsetmacro{\cy}{#2}  %
			\pgfmathsetmacro{\r}{#3}   %

			\pgfmathsetmacro{\xA}{\cx-\r}
			\pgfmathsetmacro{\yA}{\cy}
			\pgfmathsetmacro{\xB}{\cx + \r}
			\pgfmathsetmacro{\yB}{\cy}
			\pgfmathsetmacro{\nx}{\cx}
			\pgfmathsetmacro{\ny}{\cy+\r}
			\pgfmathsetmacro{\nx}{\cx}
			\pgfmathsetmacro{\ny}{\cy+\r}
			\pgfmathsetmacro{\sx}{\cx}
			\pgfmathsetmacro{\sy}{\cy-\r}

			\node at (\xA,\yA) [left] {1};
			\node at (\xB,\yB) [right] {3};
			\draw[thick] (\xA,\yA) -- (\cx,\cy) -- (\xB,\yB);
			\draw[thick] (\xA,\yA) -- (\nx,\ny) -- (\xB,\yB) --(\sx,\sy) -- (\xA,\yA);
			\fill (\xA,\yA) circle (2pt);
			\fill (\xB,\yB) circle (2pt);
			\fill (\nx,\ny) circle (2pt);
			\fill (\sx,\sy) circle (2pt);
			\draw[thick, black] (\xA,\yA) to[out=45, in=90+45] (\xB,\yB);
			\draw[thick, black] (\xA,\yA) to[out=270+45, in=270-45] (\xB,\yB);
			\draw[thick, black] (\cx,\cy) to[out=45, in=90+45] (\xB,\yB);
			\draw[thick, black] (\cx,\cy) to[out=270+45, in=270-45] (\xB,\yB);
			\node at (0,\r) [above] {2};
			\node at (0,-\r) [below] {4};
			\fill[red] (\cx,\cy) circle (2pt);
			\fill[blue] ($0.5*(\xB,\yB)$) circle (2pt);
		}
		\chamfourtadpole{0}{0}{1.2}
	\end{tikzpicture}
	\caption{The eight labels that appear for the two-loop odd half-chambers. The red/blue vertex is associated to the loop momentum $y_1/y_2$.}
	\label{fig:4pt-2lopp-half-chambers}
\end{figure}

As an example, let us focus on the two-loop half-chambers for $n=4$. Since there is a single tree-level and one-loop chamber, we can treat the data $({\bf x};y_1)$ as fixed. The two-loop fiber contains vertices ${\bf x}$ together with the four triple-cut points $\{ q^+_{13y_1}, q^-_{13y_1},q^+_{24y_1},q^-_{24y_1}\}$. Using the null cones of the odd vertices we can split the two-loop fiber into two-loop half-chambers. We find that the two-loop fiber is split into $8$ regions whose labels are displayed in Fig. \ref{fig:4pt-2lopp-half-chambers}.

\section{Relation to BCFW}
\label{sec:BCFW}
In this section we show the connection between the half-chamber expansion and the BCFW expansion of ABJM integrands.
\subsection{Tree-level}
We focus first on tree level, where the BCFW recursion for ABJM amplitudes is represented in terms of on-shell diagrams as \cite{Gang:2010gy, Huang:2013owa}
\begin{align}
	\raisebox{-30.25pt}{\includegraphics[scale=1.5]{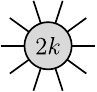}} =\sum_{k_l+k_r=k+1} &\raisebox{-27.5pt}{\includegraphics[scale=1.5]{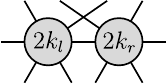}}\,.
\end{align}
Since our discussion is largely taking place in dual momentum space, it is natural to associate these diagrams to their dual, in which case the BCFW recursion takes the form
\begin{align}\label{eq:BCFW-tree}
	\raisebox{-24pt}{\includegraphics[scale=0.9]{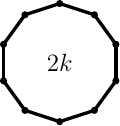}} \  \ =\sum_{k_l+k_r=k+1}  \ \raisebox{-24pt}{\includegraphics[scale=0.9]{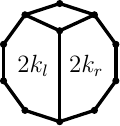}}\,.
\end{align}
The lower-point amplitudes on the right-hand side can themselves be decomposed recursively until only four-point tree amplitudes remain. We will follow a specific recursion scheme: at each step, an internal vertex is introduced into every subpolygon with $n>4$ vertices. This vertex is then connected to the two vertices adjacent to the vertex added in the previous step, as well as to one additional vertex chosen such that the resulting subpolygons each have an even number of vertices. By repeating this procedure, the recursion systematically reduces all subpolygons to four-point amplitudes. For example, at ten points the recursion takes the following form
\begin{align}
	\raisebox{-22pt}{\includegraphics[scale=0.8]{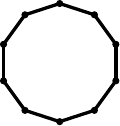}} & \ = \ \raisebox{-22pt}{\includegraphics[scale=0.8]{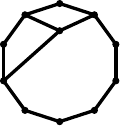}} \ + \ \raisebox{-22pt}{\includegraphics[scale=0.8]{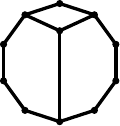}} \ +\ \raisebox{-22pt}{\includegraphics[scale=0.8]{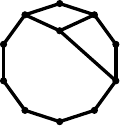}} \notag \\
	& \ =\ \raisebox{-22pt}{\includegraphics[scale=0.8]{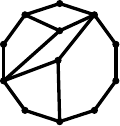}}\ +\ \raisebox{-22pt}{\includegraphics[scale=0.8]{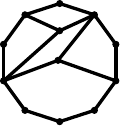}} \ +\ \raisebox{-22pt}{\includegraphics[scale=0.8]{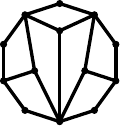}} \ +\ \raisebox{-22pt}{\includegraphics[scale=0.8]{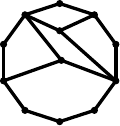}} \ +\ \raisebox{-22pt}{\includegraphics[scale=0.8]{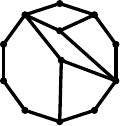}} \notag \\
	&\ =\ \raisebox{-22pt}{\includegraphics[scale=0.8]{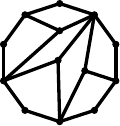}}\ +\ \raisebox{-22pt}{\includegraphics[scale=0.8]{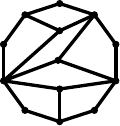}} \ +\ \raisebox{-22pt}{\includegraphics[scale=0.8]{Figures/bcfw_2_1}} \ +\ \raisebox{-22pt}{\includegraphics[scale=0.8]{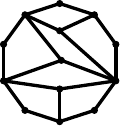}}\ +\ \raisebox{-22pt}{\includegraphics[scale=0.8]{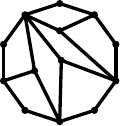}}.
\end{align}
In the dual on-shell diagram, we now consider replacing an internal vertex connected to vertices $i,j,k$ with the triangle $t_{ijk}$ through a `$Y$-$\Delta$ transform'. Our choice of BCFW recursion guarantees that the triangles share internal chords. As an example, after applying the $Y$-$\Delta$ transform to each internal vertex, the final form of the $10$-point recursion is given by
\begin{align}
	\raisebox{-22pt}{\includegraphics[scale=0.8]{Figures/bcfw_dec}} \ =\ \raisebox{-22pt}{\includegraphics[scale=0.8]{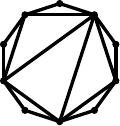}}\ +\ \raisebox{-22pt}{\includegraphics[scale=0.8]{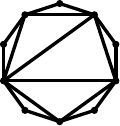}}\ +\ \raisebox{-22pt}{\includegraphics[scale=0.8]{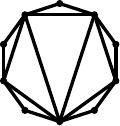}}\ +\ \raisebox{-22pt}{\includegraphics[scale=0.8]{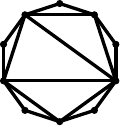}}\ +\ \raisebox{-22pt}{\includegraphics[scale=0.8]{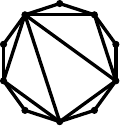}}.
\end{align}
Written in this form, the BCFW recursion can be interpreted as a sum over triangulations of the all-even or all-odd subpolygon, depending on the initial choice of legs used in the recursion. Diagrammatically, this is equivalent to the Berends-Giele recursion for tree-level $k$-point amplitudes in $\Tr(\phi^3)$, that is 
\begin{align}
	\raisebox{-25pt}{\includegraphics[scale=0.9]{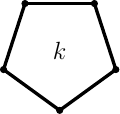}} \  \ =\displaystyle\text{{\Large$\displaystyle\sum_{\begin{smallmatrix} k_l+k_r=k+1\end{smallmatrix}}$}}  \ \ \raisebox{-25pt}{\includegraphics[scale=0.9]{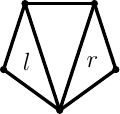}}.
\end{align}
To convert a triangulation of the odd/even subpolygon back into a traditional on-shell diagram we simply take the {\it medial graph} of the triangulation i.e. for each edge of the triangulation we introduce a vertex and connect all vertices which appear as an edge in a triangle. An example of this procedure (omitting all external legs) for the ten-point tree-level amplitude is given by
\begin{align}
	\raisebox{-22pt}{\includegraphics[scale=0.8]{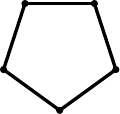}}  \ \ &= \  \ \raisebox{-20pt}{\includegraphics[scale=0.8]{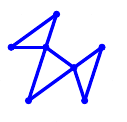}} \ + \ \raisebox{-20pt}{\includegraphics[scale=0.8]{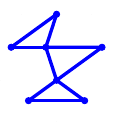}} \ + \  \raisebox{-20pt}{\includegraphics[scale=0.8]{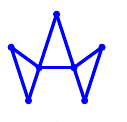}} \ + \  \raisebox{-20pt}{\includegraphics[scale=0.8]{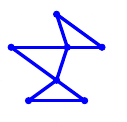}}+\ \raisebox{-20pt}{\includegraphics[scale=0.8]{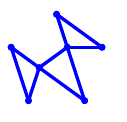}}. \notag 
\end{align}
Note, this prescription for the tree-level BCFW recursion appeared already in \cite{Huang:2013owa,Huang:2014xza}, and the corresponding on-shell diagrams were termed `trees of triangles' in \cite{Perlstein:2025lgv}.

Thus, using the diagrams we have introduced, the BCFW recursion for tree-level ABJM amplitudes is equivalent to the sum over all odd/even triangulations of an $n$-gon. But this is exactly the combinatorial data we associate to tree-level half-chambers! In fact, we claim that this relation is precise: given an odd/even triangulation of an $n$-gon, the associated on-shell function is equivalent to the canonical form of the associated half-chamber. The argument for this claim is straightforward. We first note that each term in the expansion \eqref{eq:BCFW-tree} can be interpreted as a leading singularity of a one-loop integrand $A_n^{(1)}$. Explicitly, 
\begin{align}
	\mathop{\Res}_{\substack{(y-x_i)^2=0\\(y-x_j)^2=0\\(y-x_k)^2=0}} A_n^{(1)}=\;\vcenter{\hbox{\includegraphics[height=25mm]{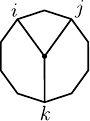}}}\,.
\end{align}
We know that we can expand the one-loop integrand in terms of tree-level chambers and one-loop fibers (\textit{cf.} equation \eqref{eq:one-loop-chamber-fib}). By definition, the canonical form of $\Delta({\bf x})$ satisfies that 
\begin{align}
	\mathop{\Res}_{\substack{(y-x_i)^2=0\\(y-x_j)^2=0\\(y-x_k)^2=0}} \Omega[\Delta({\bf x})] = \begin{cases}
		1 \quad &\text{if }q^\pm_{ijk}\text{ a vertex of }\Delta({\bf x})\\
		0 &\text{if }q^\pm_{ijk}\text{ not a vertex of }\Delta({\bf x})
	\end{cases}.
\end{align}
It is therefore clear that this residue is only supported by terms where the one-loop fiber $\Delta({\bf x})$ contains the vertex $q^\pm_{ijk}$. Furthermore, the result of the residue of $A_n^{(1)}$ will therefore be the sum of canonical forms of all tree-level chambers which contain the triangle $t_{ijk}$ in their triangulation. Similarly, we find that the on-shell functions labelled by any (partial) odd/even triangulation of an $n$-gon is given exactly by the sum of chambers which contain all the triangles in the triangulation. We thus see that the on-shell function of a full odd/even triangulation is exactly the canonical form of the tree-level half-chamber with the same label.

As we have repeatedly emphasised throughout this paper, the dual of an odd/even triangulation of an $n$-gon is a tree-level Feynman diagram for $k$-point $\Tr(\phi^3)$, and the BCFW cells are thus in a bijection with these Feynman diagrams. For Feynman diagrams, the internal edges correspond to physical poles of the amplitude, but for the corresponding BCFW cell these edges correspond to spurious boundaries instead. Said differently, two BCFW cells share a spurious boundary when they are related through an `$s$-$t$ channel flip'. Since the boundary structure of $k$-point tree-level amplitudes in $\Tr(\phi^3)$ is captured by the associahedron \cite{Arkani-Hamed:2017mur}, we see that the \emph{adjacency graph} of BCFW cells for ABJM is also an associahedron.

\subsection{Loop-level}
The all-loop extension of the BCFW recursion for integrands in ABJM reads \cite{Huang:2014xza}
\begin{align}
	\raisebox{-30.25pt}{\includegraphics[scale=1.5]{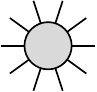}} =\sum_{L,R} &\raisebox{-27.5pt}{\includegraphics[scale=1.5]{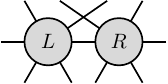}}+ \raisebox{-31pt}{\includegraphics[scale=1.5]{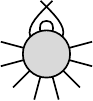}}\,.
\end{align}
Using the same logic as for the tree-level case, we find that the all-loop recursion can be represented graphically as a recursion enumerating all triangulations (without tadpoles) of an $L$-punctured $k$-gon as\footnote{We thank Hadleigh Frost for introducing us to this recursion relation and for providing a generating function for the number of triangulations (with tadpoles) of an $L$-punctured $k$-gon.}  
\begin{align}
	\raisebox{-25pt}{\includegraphics[scale=0.9]{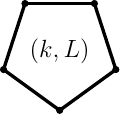}} \ =\displaystyle\text{{\Large$\displaystyle\sum_{\begin{smallmatrix} k_l+k_r=k+1,\\L_l+L_r=L\end{smallmatrix}}$}}  \ \raisebox{-25pt}{\includegraphics[scale=0.9]{Figures/bcfw_tree_right}} \ + \ \raisebox{-25pt}{\includegraphics[scale=0.9]{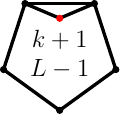}}.
\end{align}
Again, this recursion is diagrammatically equivalent to the Berends-Giele recursion for $L$-loop $\Tr(\phi^3)$ integrands.  As a further example at six-point one-loop we have two factorisation and one forward-limit term resulting in the following
\begin{align}
	\raisebox{-17pt}{\includegraphics[scale=0.7]{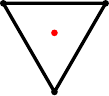}}\! &=\raisebox{-17pt}{\includegraphics[scale=0.7]{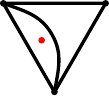}}\ +\ \raisebox{-17pt}{\includegraphics[scale=0.7]{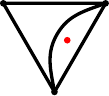}}\ +\ \raisebox{-17pt}{\includegraphics[scale=0.7]{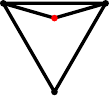}} \notag \\ 
	&=\raisebox{-17pt}{\includegraphics[scale=0.7]{Figures/three_point_one_loop_fac1}}\ +\ \raisebox{-17pt}{\includegraphics[scale=0.7]{Figures/three_point_one_loop_fac2}}\ +\ \raisebox{-17pt}{\includegraphics[scale=0.7]{Figures/three_point_one_loop_fl1}} \ +\ \raisebox{-17pt}{\includegraphics[scale=0.7]{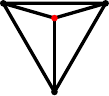}}.
\end{align}
By taking the medial graph of these triangulations we arrive at the familiar representations of the 6-point one-loop integrand in terms of on-shell diagrams given by 
\begin{align}
	\raisebox{-17pt}{\includegraphics[scale=0.7]{Figures/three_point_one_loop_lhs}}\! &= \ \raisebox{-15pt}{\includegraphics[scale=1.1]{Figures/bcfw_example_6_pts_2}}  \ +\  \raisebox{-15pt}{\includegraphics[scale=1.1]{Figures/bcfw_example_6_pts_3}} \ +\ \raisebox{-15pt}{\includegraphics[scale=1.1]{Figures/bcfw_example_6_pts_1}} \ +\  \raisebox{-10pt}{\includegraphics[scale=1.1]{Figures/bcfw_example_6_pts}}.
\end{align}
Remarkably, as noted in section \ref{sec:half-chambers}, the $L$-loop half-chambers are captured by the very same diagrams. We again claim that the canonical form of the half-chamber is equivalent to the corresponding on-shell function. Hence, the half-chamber triangulation of the ABJM momentum amplituhedron gives a precise geometric interpretation to the loop-level BCFW expansion. The notion of compatible half-chambers which define an $L$-loop chamber can thus equivalently be phrased as a question about `compatibility' between BCFW cells.

Performing the recursion relation we find that the number of terms in the BCFW representation for the $L$-loop integrand are counted by the number of triangulations (without tadpoles) of an $L$-punctured $k$-gon given explicitly by\footnote{We have introduced the Pochhammer symbol $(x)_{a+1}=x(x+1)\ldots(x+a)$. }
\begin{align}\label{eq:nBCFW}
	n_{\text{BCFW}}(k,L)&=\frac{2^{ L}}{L!} \binom{2k-2}{k-1} {(k-1) \left(2L+2k-1 \right)_{L-2} }\,.
\end{align}
This formula extends to all loops the counting already observed at tree-level, one-loop, and two-loops in \cite{Huang:2013owa,Huang:2014xza}, where it reduces to the Catalan numbers, $\binom{2 k - 2}{ k - 2}$ and $4 (k - 1) \binom{2 k - 3}{k-2}$ respectively. Note, the $1/L!$ factor takes into account the symmetrisation of each diagram with respect to the loop variables.

\section{Loop chambers}\label{sec:loop-chambers}

So far we have mainly considered $L$-loop \emph{half} chambers: regions of $\mathcal{O}_k^{(L)}$ which determine \emph{half} of the vertices of the $(L+1)$-loop fiber. The \emph{full} structure of the $(L+1)$-loop fiber is captured by $L$-loop chambers, which arise as the intersection of one odd and one even $L$-loop half-chamber. 

As we reviewed in section \ref{sec:tree-level-chambers}, the tree-level half-chambers are given by odd/even triangulations of an $n$-gon, and the tree-level chambers are given by pairs of such triangulations. Any combination is allowed: an odd and an even tree-level half-chamber will always intersect. At loop level, the half-chambers are given by triangulations of an $L$-punctured $k$-gon, or, dually, by an $L$-loop Feynman diagram. Again, the $L$-loop chambers are given by pairs of these labels, however it is no longer the case that \emph{any} pair of odd/even half-chambers will intersect. Out of the $n_{\text{BCFW}}(k,L)^2$ possible $L$-loop chambers only some subset will have a non-empty intersection.

We recall from section \ref{sec:generaln} that we can associate to each one-loop half-chamber a corresponding sub-polygon of an $n$-gon. A necessary condition for two half-chambers to intersect is that the associated sub-polygons overlap. However, this is not a sufficient condition and a full classification of which half-chambers intersect is currently not known beyond the examples we present in this paper, i.e. four-point one and two loops, six-point one loop, and eight-point one loop. 

\subsection{Examples}
The remainder of this section will be used to study several examples of loop chambers. 

\subsubsection{Four-point one-loop}\label{sec:n4-L1-chambers}
\begin{figure}
	\centering
	\begin{tikzpicture}

		\newcommand{\chamfourtadpole}[3]{
			\pgfmathsetmacro{\cx}{#1}  %
			\pgfmathsetmacro{\cy}{#2}  %
			\pgfmathsetmacro{\r}{#3}   %

			\pgfmathsetmacro{\xA}{\cx-\r}
			\pgfmathsetmacro{\yA}{\cy}
			\pgfmathsetmacro{\xB}{\cx + \r}
			\pgfmathsetmacro{\yB}{\cy}
			\pgfmathsetmacro{\nx}{\cx}
			\pgfmathsetmacro{\ny}{\cy+\r}
			\pgfmathsetmacro{\nx}{\cx}
			\pgfmathsetmacro{\ny}{\cy+\r}
			\pgfmathsetmacro{\sx}{\cx}
			\pgfmathsetmacro{\sy}{\cy-\r}

			\node at (\xA,\yA) [left] {1};
			\node at (\xB,\yB) [right] {3};
			\draw[thick] (\xA,\yA) -- (\cx,\cy) -- (\xB,\yB);
			\draw[thick] (\xA,\yA) -- (\nx,\ny) -- (\xB,\yB) --(\sx,\sy) -- (\xA,\yA);
			\fill (\xA,\yA) circle (2pt);
			\fill (\xB,\yB) circle (2pt);
			\fill (\nx,\ny) circle (2pt);
			\fill (\sx,\sy) circle (2pt);
			\draw[thick, black] (\xA,\yA) to[out=45, in=90+45] (\xB,\yB);
			\draw[thick, black] (\xA,\yA) to[out=270+45, in=270-45] (\xB,\yB);
			\node at (0,\r) [above] {2};
			\node at (0,-\r) [below] {4};
			\fill[red] (\cx,\cy) circle (2pt);
		}
		\chamfourtadpole{0}{0}{1.4}
	\end{tikzpicture}
	\begin{tikzpicture}

		\newcommand{\chamfourtadpole}[3]{
			\pgfmathsetmacro{\cx}{#1}  %
			\pgfmathsetmacro{\cy}{#2}  %
			\pgfmathsetmacro{\r}{#3}   %

			\pgfmathsetmacro{\xA}{\cx-\r}
			\pgfmathsetmacro{\yA}{\cy}
			\pgfmathsetmacro{\xB}{\cx + \r}
			\pgfmathsetmacro{\yB}{\cy}
			\pgfmathsetmacro{\nx}{\cx}
			\pgfmathsetmacro{\ny}{\cy+\r}
			\pgfmathsetmacro{\nx}{\cx}
			\pgfmathsetmacro{\ny}{\cy+\r}
			\pgfmathsetmacro{\sx}{\cx}
			\pgfmathsetmacro{\sy}{\cy-\r}

			\node at (\xA,\yA) [left] {1};
			\node at (\xB,\yB) [right] {3};
			\draw[thick] (\nx,\ny) -- (\sx,\sy);
			\draw[thick] (\xA,\yA) -- (\nx,\ny) -- (\xB,\yB) --(\sx,\sy) -- (\xA,\yA);
			\fill (\xA,\yA) circle (2pt);
			\fill (\xB,\yB) circle (2pt);
			\fill (\nx,\ny) circle (2pt);
			\fill (\sx,\sy) circle (2pt);
			\draw[thick, black] (\nx,\ny) to[out=270+45, in=90-45] (\sx,\sy);
			\draw[thick, black] (\nx,\ny) to[out=270-45, in=90+45] (\sx,\sy);
			\node at (0,\r) [above] {2};
			\node at (0,-\r) [below] {4};
			\fill[red] (\cx,\cy) circle (2pt);
			
		}
		\chamfourtadpole{0}{0}{1.4}
	\end{tikzpicture}
	\quad \quad \quad  \quad
	\begin{tikzpicture}

		\newcommand{\chamfourtadpole}[3]{
			\pgfmathsetmacro{\cx}{#1}  
			\pgfmathsetmacro{\cy}{#2}  
			\pgfmathsetmacro{\r}{#3}

			\pgfmathsetmacro{\xA}{\cx-\r}
			\pgfmathsetmacro{\yA}{\cy}
			\pgfmathsetmacro{\xB}{\cx + \r}
			\pgfmathsetmacro{\yB}{\cy}
			\pgfmathsetmacro{\nx}{\cx}
			\pgfmathsetmacro{\ny}{\cy+\r}
			\pgfmathsetmacro{\nx}{\cx}
			\pgfmathsetmacro{\ny}{\cy+\r}
			\pgfmathsetmacro{\sx}{\cx}
			\pgfmathsetmacro{\sy}{\cy-\r}
			
			\node at (0,\r) [above] {};
			\node at ($1.2*(0,-\r)$) [below] {};
			\fill[blue] ($0.6*(\nx,\ny)$) circle (2pt);
			\fill[blue] ($0.6*(\sx,\sy)$) circle (2pt);
			\fill[blue] ($0.2*(\nx,\ny)$) circle (2pt);
			\fill[blue] ($0.2*(\sx,\sy)$) circle (2pt);
			\draw[thick,blue] ($0.6*(\nx,\ny)$) -- ($0.2*(\nx,\ny)$);
			\draw[thick,blue] ($0.6*(\sx,\sy)$) -- ($0.2*(\sx,\sy)$);
			\draw[thick, blue] ($0.2*(\nx,\ny)$) to[out=220, in=150] ($0.2*(\sx,\sy)$);
			\draw[thick, blue] ($0.2*(\nx,\ny)$) to[out=330, in=30] ($0.2*(\sx,\sy)$);
			\node at ($0.6*(\nx,\ny)$) [above,blue] {$x_2$};
			\node at ($0.2*(\nx,\ny)$) [left,blue] {$q^-_{13y}$};
			\node at ($0.2*(\sx,\sy)$) [left,blue] {$q^+_{13y}$};
			\node at ($0.6*(\sx,\sy)$) [below,blue] {$x_4$};
		}
		\chamfourtadpole{0}{0}{1.4}
	\end{tikzpicture}
	\quad
	\begin{tikzpicture}

		\newcommand{\chamfourtadpole}[3]{
			\pgfmathsetmacro{\cx}{#1}  
			\pgfmathsetmacro{\cy}{#2}  
			\pgfmathsetmacro{\r}{#3}   

			\pgfmathsetmacro{\xA}{\cx-\r}
			\pgfmathsetmacro{\yA}{\cy}
			\pgfmathsetmacro{\xB}{\cx + \r}
			\pgfmathsetmacro{\yB}{\cy}
			\pgfmathsetmacro{\nx}{\cx}
			\pgfmathsetmacro{\ny}{\cy+\r}
			\pgfmathsetmacro{\nx}{\cx}
			\pgfmathsetmacro{\ny}{\cy+\r}
			\pgfmathsetmacro{\sx}{\cx}
			\pgfmathsetmacro{\sy}{\cy-\r}
			
			\node at (0,\r) [above] {};
			\node at ($1.2*(0,-\r)$) [below] {};
			\fill[blue] ($0.6*(\xA,\yA)$) circle (2pt);
			\fill[blue] ($0.6*(\xB,\yB)$) circle (2pt);
			\fill[blue] ($0.2*(\xA,\yA)$) circle (2pt);
			\fill[blue] ($0.2*(\xB,\yB)$) circle (2pt);
			\draw[thick,blue] ($0.6*(\xA,\yA)$) -- ($0.2*(\xA,\yA)$);
			\draw[thick,blue] ($0.6*(\xB,\yB)$) -- ($0.2*(\xB,\yB)$);
			\draw[thick, blue] ($0.2*(\xA,\yA)$) to[out=0+60, in=180-60] ($0.2*(\xB,\yB)$);
			\draw[thick, blue] ($0.2*(\xA,\yA)$) to[out=0-60, in=180+60] ($0.2*(\xB,\yB)$);
			\node at ($0.6*(\xA,\yA)$) [left,blue] {$x_1$};
			\node at ($0.6*(\xB,\yB)$) [right,blue] {$x_3$};
			\node at ($0.2*(\xA,\yA)+(0,0.05)$) [above,blue] {$q^+_{24y}$};
			\node at ($0.2*(\xB,\yB)-(0,0.05)$) [below,blue] {$q^-_{24y}$};
		}
		\chamfourtadpole{0}{0}{1.4}
	\end{tikzpicture}
	\caption{The pair of triangulations corresponding to the single one-loop chamber for $n=4$.}
	\label{fig:four_point_one_loop_chams}
\end{figure}
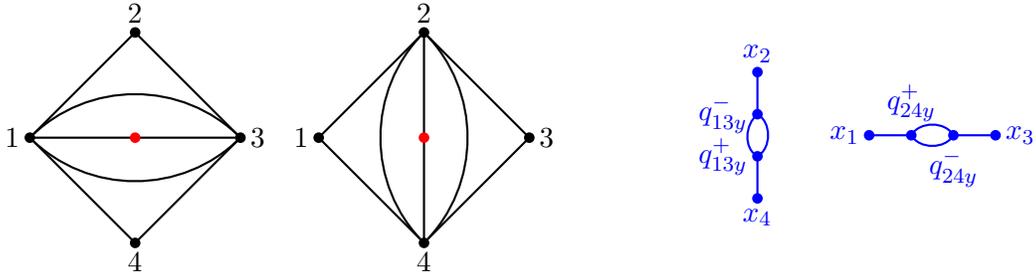

\begin{figure}[h]
	\centering
	\includegraphics[width=0.4\textwidth]{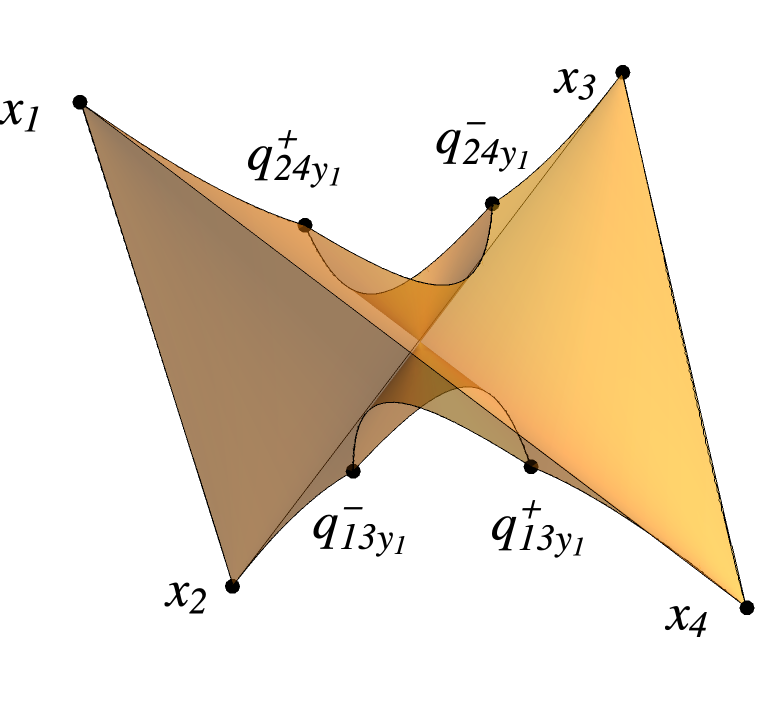} 
	\caption{The representative of the two-loop fiber for $n=4$.}
	\label{fig:4pt-2loop-fiber} 
\end{figure}

At four-points there is a single odd and a single even half-chamber, resulting in a single one-loop chamber whose label is depicted on the left of Fig. \ref{fig:four_point_one_loop_chams}. The dual of this triangulation is depicted in blue on the same figure. Comparing this to the two-loop fiber geometry depicted in Fig. \ref{fig:4pt-2loop-fiber} we conclude that (the dual of) the one-loop chamber labels completely specify the one-skeleton of the corresponding two-loop fiber. This is analogous to the situation at one-loop where (the dual of) the labels for the tree-level chamber completely specify the one-skeleton of the corresponding one-loop fiber.

\subsubsection{Six-point one-loop}
In section \ref{sec:six-point-half-chambers} we saw that there are four odd and four even six-point one-loop half-chambers. As was already observed in \cite{Lukowski:2023nnf}, only $13$ out of the $16$ possible intersections are non-empty. The compatibility of the half-chambers is summarised in Table \ref{tab:chams_6_text}.
\begin{table}[h!]
	\centering
	\begin{tabular}{m{2.5cm}||c|c|c|c}
		& \includegraphics[scale=0.8]{Figures/six_point_hc_4} &\includegraphics[scale=0.8]{Figures/six_point_hc_1} & \includegraphics[scale=0.8]{Figures/six_point_hc_2} & \includegraphics[scale=0.8]{Figures/six_point_hc_3} \\
		\hline\hline
		\includegraphics[scale=0.8]{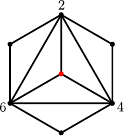}&\color{green}{{\huge \cmark}} &\color{green}{{\huge \cmark}} &\color{green}{{\huge \cmark}}&\color{green}{{\huge \cmark}}\\
		\hline
			\includegraphics[scale=0.8]{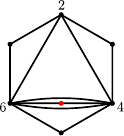}&\color{green}{{\huge \cmark}} & \color{red}{{\huge \xmark}} &\color{green}{{\huge \cmark}} & \color{green}{{\huge \cmark}}\\
		\hline
					\includegraphics[scale=0.8]{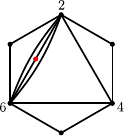} &\color{green}{{\huge \cmark}} &\color{green}{{\huge \cmark}} & \color{red}{{\huge \xmark}} &\color{green}{{\huge \cmark}}\\
		\hline
			\includegraphics[scale=0.8]{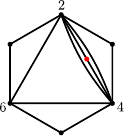}&\color{green}{{\huge \cmark}}&\color{green}{{\huge \cmark}}&\color{green}{{\huge \cmark}}&\color{red}{{\huge \xmark}}
	\end{tabular}
	\caption{The thirteen one-loop chambers for $n=6$ labelled by pairs of triangulations of a punctured $3$-gon. The green check marks indicate which half-chambers are compatible.}
	\label{tab:chams_6_text}
\end{table}

Let us show explicitly how the one-loop fiber $\Delta({\bf x})$ is divided into one-loop chambers. The $13$ one-loop chambers come in four different cyclic classes. Since $\mathcal{O}_3^{(1)}$ factorises as $\mathcal{O}_3^{(1)}=\mathcal{O}_3^{\text{tree}}\times\Delta({\bf x})$, we can consistently strip off the tree-level geometry and study the one-loop chambers as subsets of the one-loop fiber instead. To describe these one-loop chambers, we introduce six new points 
\begin{align}
	b_i=\mathcal{N}_{x_i}\cap\mathcal{N}_{q^+_{135}}\cap\mathcal{N}_{q^-_{135}}\,.
\end{align}
\begin{figure}
	\centering
	\includegraphics[width=\textwidth]{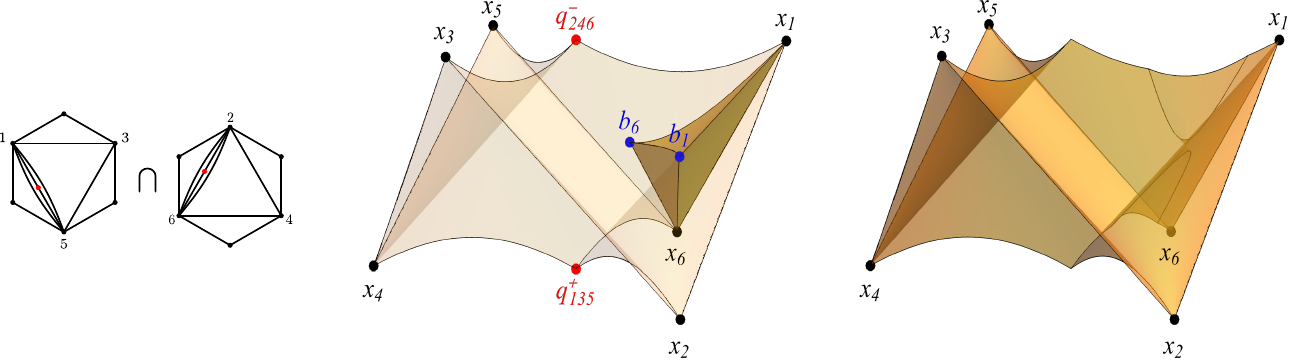}
	\caption{The six-point one-loop chamber $\mathfrak{c}_6^+=\Delta_{\widebar{1}3\widebar{5}}\cap \Delta_{\widebar{2}4\widebar{6}}$ in the middle, and the associated two-loop fiber on the right}
	\label{fig:6pt-chamber-1}
\end{figure}
The first type of one-loop chambers are of the form $\mathfrak{c}_i^+\coloneqq\Delta_{\widebar{i}\widebar{i+2}i+4}\cap \Delta_{\widebar{i-1}\widebar{i+1}i+3}$ for $i=1,\ldots,6$, the $i=6$ case is depicted in Fig. \ref{fig:6pt-chamber-1}. These chambers have vertices $x_{i-1},x_i,b_{i-1},b_i$, and their canonical form is given by
\begin{align}
	\Omega[\mathfrak{c}_i^+] = \dd\log\frac{(y-x_i)^2}{(y-q^-_{246})^2}\wedge\dd\log\frac{(y-x_{i+1})^2}{(y-q^-_{246})^2}\wedge\dd\log\frac{(y-q^+_{135})^2}{(y-q^-_{246})^2}\,.
\end{align}
\begin{figure}
	\centering
	\includegraphics[width=\textwidth]{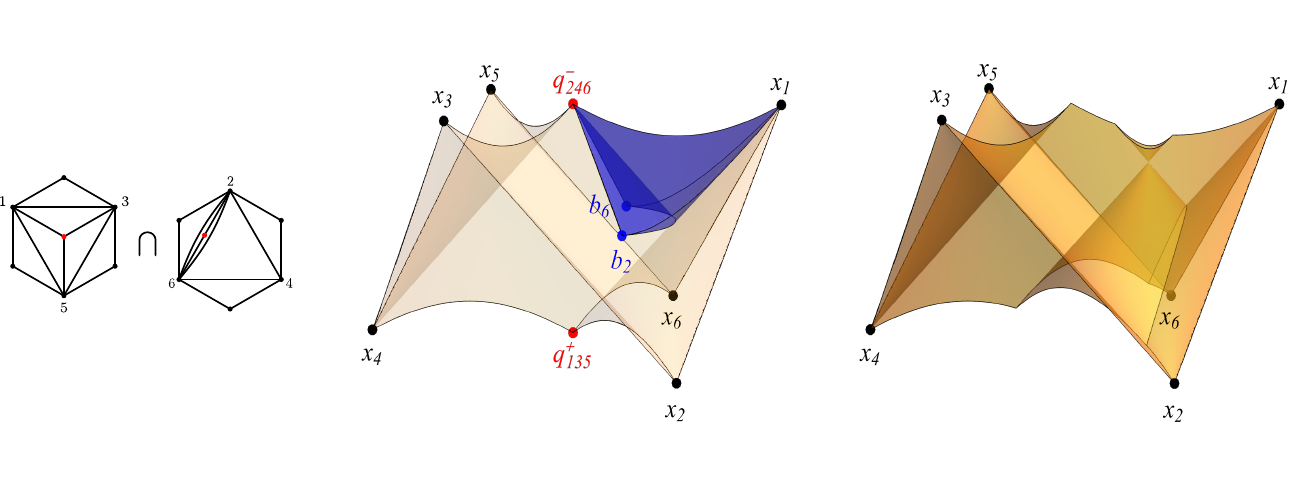}
	\caption{The six-point one-loop chamber $\mathfrak{c}_1^0=\Delta_{135}^-\cap \Delta_{\widebar{2}4\widebar{6}}$ in the middle, and the associated two-loop fiber on the right}
	\label{fig:6pt-chamber-2}
\end{figure}
The second cyclic class is of the form $\mathfrak{c}_i^0\coloneqq \Delta_{135}^-\cap\Delta_{\widebar{i-1}\widebar{i+1}i+3}$ for $i=1,3,5$, and they have vertices $x_i,b_{i-1},b_{i+1},q^-_{246}$, the case $i=1$ is depicted in Fig. \ref{fig:6pt-chamber-2}. This one-loop chamber has a canonical form
\begin{align}
	\Omega\left[\mathfrak{c}_i^0\right]_{i=1,3,5} = -\dd\log\frac{(y-x_{i-1})^2}{(y-q^-_{246})^2}\wedge\dd\log\frac{(y-x_{i+1})^2}{(y-q^-_{246})^2}\wedge\dd\log\frac{(y-x_i)^2}{(y-q^+_{135})^2}\,.
\end{align}
\begin{figure}
	\centering
	\includegraphics[width=\textwidth]{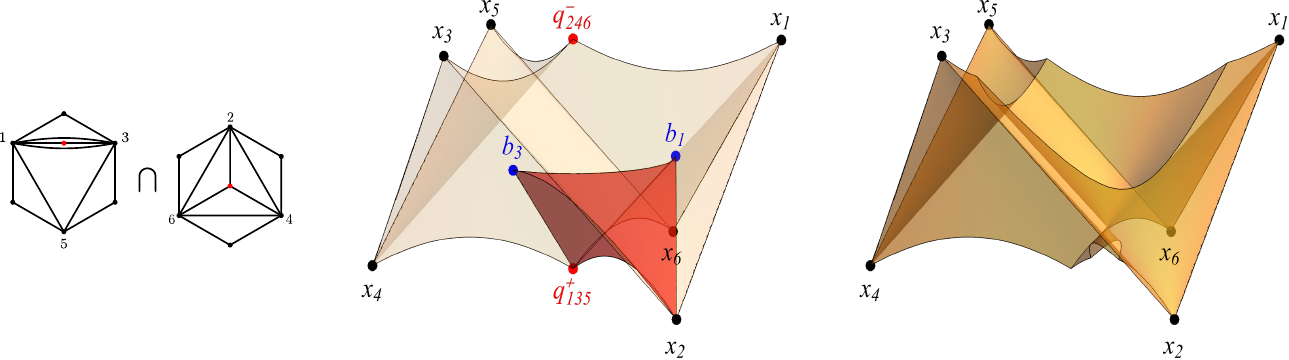}
	\caption{The six-point one-loop chamber $\mathfrak{c}_2^0=\Delta_{\widebar{1}\widebar{3}5}\cap \Delta_{246}^-$ in the middle, and the associated two-loop fiber on the right}
	\label{fig:6pt-chamber-3}
\end{figure}
Next, we have a similar type of one-loop chamber: $\mathfrak{c}_i^0=\Delta_{\widebar{i-1}\widebar{i+1}i+3}\cap\Delta_{246}^-$, $i=2,4,6,$ depicted for the case $i=2$ in Fig. \ref{fig:6pt-chamber-3}. It has vertices $x_i,b_{i-1},b_{i+1},q^+_{135}$, and a canonical form
\begin{align}
	\Omega\left[\mathfrak{c}_i^0\right]_{i=2,4,6} = \dd\log\frac{(y-x_{i-1})^2}{(y-q^+_{135})^2}\wedge\dd\log\frac{(y-x_{i+1})^2}{(y-q^+_{135})^2}\wedge\dd\log\frac{(y-x_i)^2}{(y-q^-_{246})^2}\,.
\end{align}
\begin{figure}
	\centering
	\includegraphics[width=\textwidth]{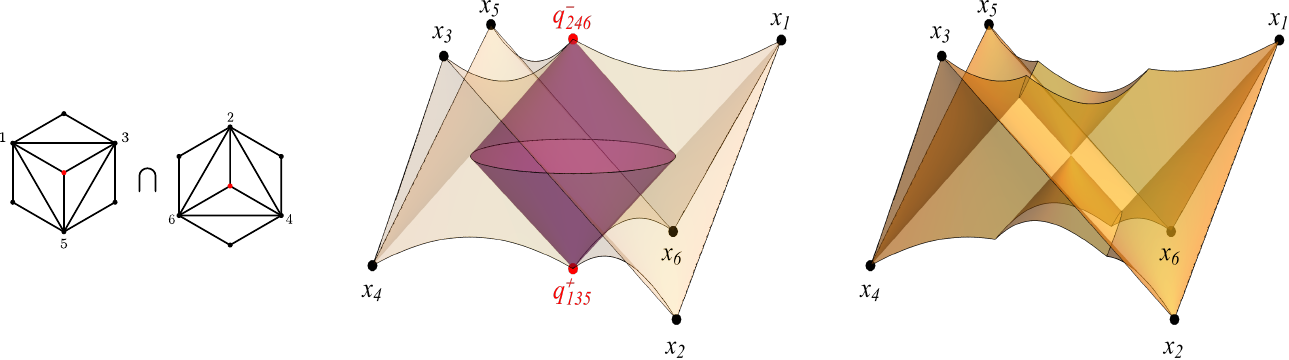}
	\caption{The six-point one-loop chamber $\mathfrak{c}^-=\Delta_{135}^-\cap \Delta_{246}^-$ in the middle, and the associated two-loop fiber on the right}
	\label{fig:6pt-chamber-4}
\end{figure}
Lastly, we have the chamber $\mathfrak{c}^-\coloneqq \Delta_{135}^-\cap\Delta_{246}^-$. This chamber is depicted in Fig. \ref{fig:6pt-chamber-4}, and does not have any vertices. As such, it also has a vanishing canonical form.

When we add up the canonical forms of the first 12 one-loop chambers, we retrieve the canonical form of the full one-loop fiber, as expected:
\begin{align}
	\Omega[\Delta({\bf x})] = \sum_{i=1}^6 \big(\Omega\left[\mathfrak{c}_i^+\right] + \Omega\left[\mathfrak{c}_i^0\right]\big)\,.
\end{align}
Although not directly obvious from the forms we present here, it is possible to find the canonical form of these lightcone geometries as a sum over vertices in the exact same way as the one-loop fibers of section \ref{sec:tree-level-chambers} and the one-loop half-chambers of section \ref{sec:six-point-half-chambers}.

\subsubsection{Four-point two-loop case}
As a two-loop example, let us consider the four-point case. Since there is a single tree-level and one-loop chamber we can treat the data $({\bf x};y_1)$ as fixed. As we noted in section \ref{sec:n4-L1-chambers}, the two-loop fiber contains vertices ${\bf x}$ together with the four triple-cut points $\{ q^+_{13y_1}, q^-_{13y_1},q^+_{24y_1},q^-_{24y_1}\}$. Using the null cones of the odd vertices we can split the two-loop fiber into two-loop half-chambers. We find that the two-loop fiber is split into $8$ regions whose labels are displayed in Fig. \ref{fig:4pt-2lopp-half-chambers}.

The full two-loop chamber consists of non-empty intersections of odd and even two-loop half-chambers. We list all the non-empty intersections in Table \ref{tab:4pt-two-loop-chambers}, where we label the half-chambers by two-loop Feynman diagrams, which are dual to the diagrams depicted in Fig. \ref{fig:4pt-2lopp-half-chambers}.
\begin{table}
	\centering
	\makebox[\textwidth][c]{
		\begin{tabular}{c||c|c|c|c|c|c|c|c}
			& \includegraphics[width=1.5cm]{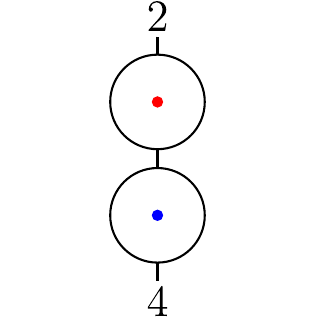} & \includegraphics[width=1.5cm]{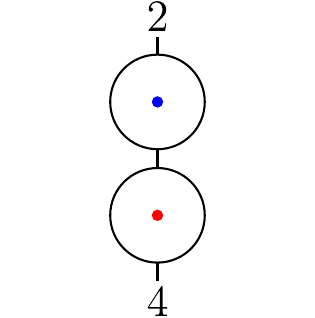} & \includegraphics[width=1.5cm]{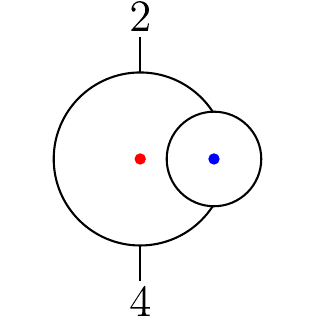} & \includegraphics[width=1.5cm]{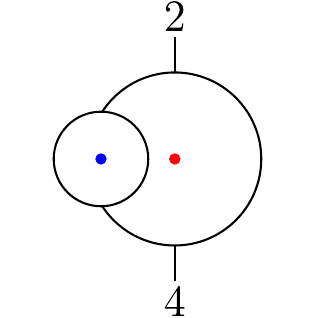} & \includegraphics[width=1.5cm]{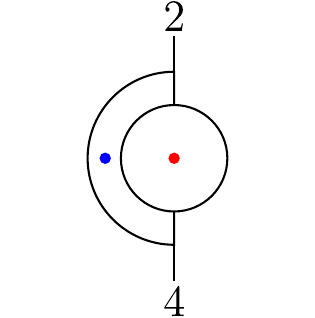} & \includegraphics[width=1.5cm]{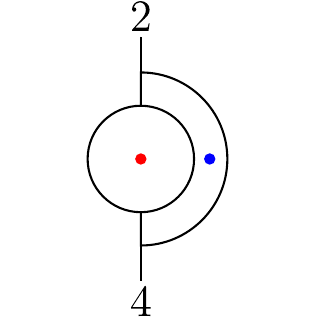} & \includegraphics[width=1.5cm]{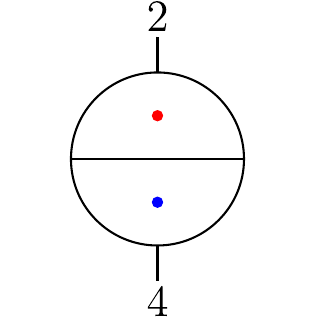} & \includegraphics[width=1.5cm]{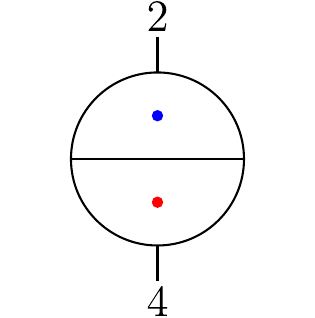} \\
			\hline\hline
			\includegraphics[width=1.5cm]{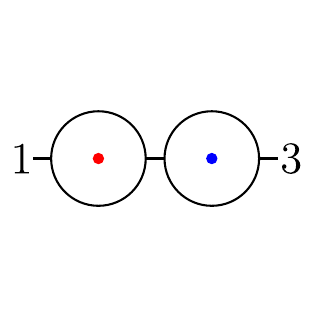} & \color{green}{{\huge \cmark}} & \color{green}{{\huge \cmark}} & \color{green}{{\huge \cmark}} & \color{red}{{\huge \xmark}} & \color{red}{{\huge \xmark}} & \color{green}{{\huge \cmark}} & \color{green}{{\huge \cmark}} & \color{green}{{\huge \cmark}} \\
			\hline
			\includegraphics[width=1.5cm]{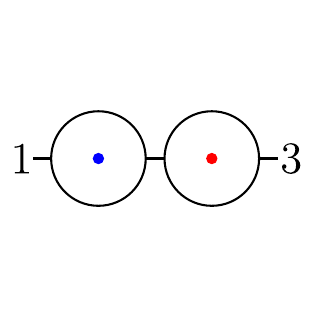} & \color{green}{{\huge \cmark}} & \color{green}{{\huge \cmark}} & \color{red}{{\huge \xmark}} & \color{green}{{\huge \cmark}} & \color{green}{{\huge \cmark}} &\color{red}{{\huge \xmark}} & \color{green}{{\huge \cmark}}  & \color{green}{{\huge \cmark}} \\
			\hline
			\includegraphics[width=1.5cm]{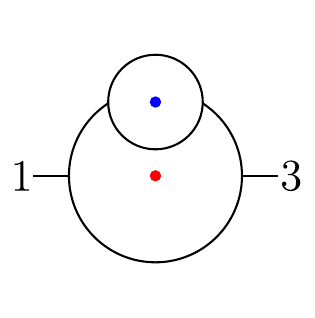} & \color{red}{{\huge \xmark}} & \color{green}{{\huge \cmark}} & \color{red}{{\huge \xmark}} & \color{red}{{\huge \xmark}} & \color{green}{{\huge \cmark}} & \color{green}{{\huge \cmark}}  & \color{red}{{\huge \xmark}} & \color{green}{{\huge \cmark}} \\
			\hline
			\includegraphics[width=1.5cm]{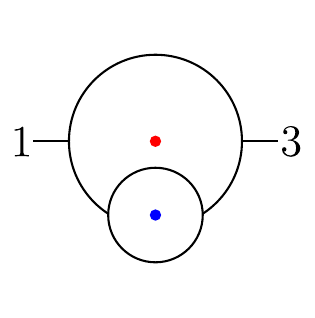} & \color{green}{{\huge \cmark}} & \color{red}{{\huge \xmark}} & \color{red}{{\huge \xmark}} & \color{red}{{\huge \xmark}} & \color{green}{{\huge \cmark}} & \color{green}{{\huge \cmark}} & \color{green}{{\huge \cmark}} & \color{red}{{\huge \xmark}} \\
			\hline
			\includegraphics[width=1.5cm]{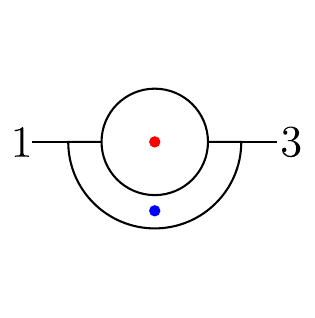} & \color{red}{{\huge \xmark}} & \color{green}{{\huge \cmark}} & \color{green}{{\huge \cmark}} & \color{green}{{\huge \cmark}}  & \color{red}{{\huge \xmark}} & \color{red}{{\huge \xmark}}  & \color{red}{{\huge \xmark}} & \color{green}{{\huge \cmark}} \\
			\hline
			\includegraphics[width=1.5cm]{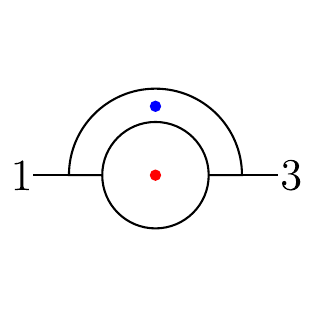} & \color{green}{{\huge \cmark}} & \color{red}{{\huge \xmark}} & \color{green}{{\huge \cmark}}  & \color{green}{{\huge \cmark}} &\color{red}{{\huge \xmark}} & \color{red}{{\huge \xmark}}  & \color{green}{{\huge \cmark}} & \color{red}{{\huge \xmark}} \\
			\hline 
			\includegraphics[width=1.5cm]{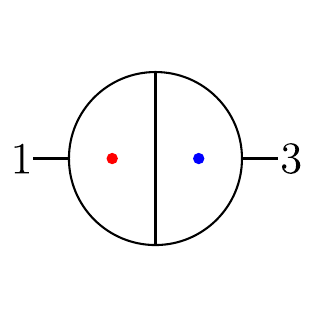} & \color{green}{{\huge \cmark}} & \color{green}{{\huge \cmark}}  & \color{green}{{\huge \cmark}} & \color{red}{{\huge \xmark}} & \color{red}{{\huge \xmark}} & \color{green}{{\huge \cmark}} & \color{green}{{\huge \cmark}} & \color{green}{{\huge \cmark}}  \\
			\hline
			\includegraphics[width=1.5cm]{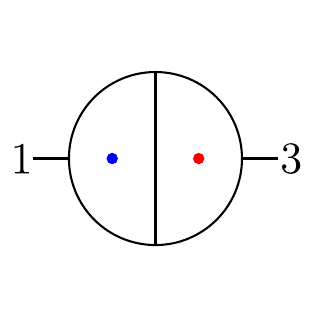} & \color{green}{{\huge \cmark}} & \color{green}{{\huge \cmark}} & \color{red}{{\huge \xmark}} & \color{green}{{\huge \cmark}} & \color{green}{{\huge \cmark}} & \color{red}{{\huge \xmark}} & \color{green}{{\huge \cmark}}  & \color{green}{{\huge \cmark}} 	 
		\end{tabular}
	}
	\caption{Table indicating which two-loop half-chambers intersect for $n=4$.}
	\label{tab:4pt-two-loop-chambers}
\end{table}

\section{Conclusions and outlook}\label{sec:conclusions}
In this paper we have investigated the geometry of loop integrands in ABJM by considering $\mathcal{O}_k^{(L+1)}$ as a fibration with a $3$-dimensional $(L+1)$-loop fiber over $\mathcal{O}_k^{(L)}$. This naturally suggests a triangulation of $\mathcal{O}_k^{(L)}$ in terms of \emph{$L$-loop chambers}. We have shown that each $L$-loop chamber can be found as the intersection of an odd and an even $L$-loop \emph{half-chamber}. These $L$-loop half-chambers inherit a natural labelling based on (half of) the combinatorial structure of the associated $(L+1)$-loop fiber. Notably, we have seen that these labels are equivalent to $L$-loop cubic Feynman diagrams with $n/2$ particles, and the set of all half-chambers is equivalent to the set of all tadpole-less Feynman diagrams. 

Furthermore, we have shown that these $L$-loop half-chambers are in a bijection with terms in the loop-level BCFW recursion. This gives a concrete geometric interpretation for the BCFW recursion. Moreover, this result highlights a curious bijection between diagrammatic expansions of loop integrands for $n$-point ABJM and $n/2$-point $\Tr(\phi^3)$ theory. The explicit form of the loop-level BCFW recursion is then equivalent to loop-level Berends-Giele recursion for $\Tr(\phi^3)$.

There are many interesting questions which naturally arise from this work. In the remainder of this section we will highlight some of the more pressing open problems.

\paragraph{Connection to $\Tr(\phi^3)$.}
The duality between BCFW cells for ABJM and Feynman diagrams in $\Tr(\phi^3)$ is highly suggestive. Physically, these two theories could hardly be more different (one a scalar theory with cubic interactions, the other a supersymmetric theory with only even-point interactions), and as such there is no a priori reason to expect any relation. It would be interesting to see if this connection can be used to find new, more concrete, formulae for loop integrands in ABJM theory. For instance, it is natural to investigate whether or not we can find some generalised current for the off-shell Berends-Giele recursion which directly calculates ABJM amplitudes.

Furthermore, there has recently been an interest in the loop integrands in $\Tr(\phi^3)$ theory and scaffolding triangulations coming from the \emph{surfaceology} framework, and the connections to amplitudes in NLSM and Yang-Mills. These novel techniques are essentially efficient ways to sum all Feynman diagrams, and hence it is natural to ask whether similar formulations can be found for ABJM.

We note that connections between ABJM theory and $\Tr(\phi^3)$ were already observed in \cite{Lukowski:2021fkf}, where it was argued that the singularity structure of tree-level ABJM amplitudes are completely contained in the singularities of $\Tr(\phi^3)$. The current discussion adds additional evidence for some connection between the two theories, although any concrete relation remains to be clarified.

\paragraph{Loop Chambers.}
We have given a complete classification for all non-empty $L$-loop half-chambers in terms of $L$-loop Feynman diagrams. Full $L$-loop chambers are given by the intersection of two `compatible' half-chambers. At tree-level, any even and odd half-chambers are compatible, giving a classification for all tree chambers as pairs of tree-level Feynman diagrams. However, we have seen that this structure does not persist to loop level, and the classification of compatible half-chambers is not known beyond the examples we worked out in this paper. 

To give a full description of the $L$-loop ABJM momentum amplituhedron by fibrating one of the loop variables, it is necessary to get a full classification of the compatible $k$-point $L$-loop Feynman diagrams (or, dually, triangulations of $k$-gons with $L$ punctures), which describe the skeletal structure of the $(L+1)$-loop fiber. Given the geometric nature from which this question arises, we expect there to be an elegant combinatorial answer. Furthermore, due to the relevance of these diagrams for other areas of physics and mathematics, it would be interesting to see if this notion of compatibility is of relevance in other contexts.

\paragraph{Breakdown of Fibrations.}
The idea of fibrating a single loop variable over a lower-loop object can be iterated to write $\mathcal{O}_k^{(L)}$ as a sum over tree chambers times $L$ three-dimensional lightcone geometries. This is equivalent to the approach taken in \cite{Ferro:2024vwn} for loop integrands in $\mathcal{N}=4$ SYM. Unfortunately, for the case at hand the canonical form no longer distributes over this product. To highlight this point, let us look at a simple 2-loop example for $n=4$. We have seen that there is only a single one-loop chamber, $\mathcal{O}_2^{(1)}$, and hence the geometry splits up like
\begin{align}
	\mathcal{O}_2^{(2)}=\mathcal{O}_2^{(1)}\times f^{(2)}\,,
\end{align}
where $f^{(2)}$ is the two-loop fiber depicted in Fig. \ref{fig:4pt-2loop-fiber}. This is a valid factorisation of the geometry, however it does not extend to the level of canonical forms:
\begin{align}
	\Omega[\mathcal{O}_2^{(2)}]\neq \Omega[\mathcal{O}_2^{(1)}]\wedge \Omega[f^{(2)}]\,.
\end{align}
This mismatch between the factorisation of the geometry and the canonical form is somewhat surprising and goes contrary to the usual `chambers and fibrations' philosophy. At this stage it is not quite clear where this important property breaks down, especially as it is known to work in many other cases \cite{He:2023rou,Lukowski:2023nnf,Ferro:2023qdp,Ferro:2024vwn,Glew:2024zoh,He:2025rza,He:2024xed,Bartsch:2025mvy}. 

It is important to investigate when this logic breaks down. Gaining a better understanding of these cases might help remedy the above discrepancy. This is particularly important since the canonical form of lightcone geometries can easily be found once the vertex structure is known. Hence, if there is a way the above discrepancy can be remedied, the combinatorial information on loop chambers and fibers we uncovered in this paper could directly be used to find efficient new formulae for higher loop integrands for ABJM.

\paragraph{Extension to $\mathcal{N}=4$ SYM.}
Scattering amplitudes in ABJM theory are remarkably similar in structure to those of $\mathcal{N}=4$ SYM. Of particular importance is the fact that $\mathcal{N}=4$ SYM also allows a geometric description in terms of loop fibers which consists of lightcone geometries in dual space \cite{Ferro:2023qdp}. Hence, it is natural to ask whether the techniques and results of this paper can be generalised to this setting. Particularly, it would be interesting to see if the study of $L$-loop chambers for $\mathcal{N}=4$ SYM can be used to rediscover the loop-level BCFW recursion. This would provide an alternative geometric interpretation to loop level BCFW.

\paragraph{Negative Geometries.}
An object of recent study both in $\mathcal{N}=4$ SYM and ABJM theory is the so-called {\it negative geometry} first introduced in \cite{Arkani-Hamed:2021iya}. The negative geometries arise from the definition of the amplituhedron upon relaxing or flipping the sign of the mutual positivity conditions between loop momenta. The simplest class of such geometries are given by the all negative ladder geometries characterised by the constraints $(y_a-y_{a+1})^2<0$ for all $a \in \{1,\ldots,L-1\}$. The negative ladder geometries for MHV integrands in $\mathcal{N}=4$ SYM have recently been computed for all $n$ and $L$ in \cite{Glew:2024zoh}, and it would be interesting to see whether the analogous result can also be obtained for ABJM theory. We expect the chambers introduced in this paper to aid in this calculation.

Furthermore, we have shown that fibrating a loop variable and splitting the lightcone of $y_L$ into a future and a past lightcone naturally leads to the discovery of the loop-level BCFW triangulation. These ideas can easily be applied for negative geometries as well, and as such it would be interesting to see if we can use this logic to derive BCFW-like recursion for negative geometries.

\paragraph{BDS Integrand for ABJM.}
Very recently, the scaffolding triangulations which describe tree-level BCFW have proven to be useful for the description of the so-called BDS integrand of ABJM \cite{Huang:2025hya}. It was further shown that this graphical representation remains in tact after integration. The work we presented in this paper gives a natural generalisation of these diagrammatics for higher loop BCFW, and as such it would be interesting to see whether this can be given a similar interpretation in relation to the BDS integrand.

\section*{Acknowledgements}
We thank the Galileo Galilei Institute for Theoretical Physics for the hospitality and the INFN for partial support during the completion of this work. This
work was supported in part by the Deutsche Forschungsgemeinschaft (DFG, German Research Foundation) Projektnummer 508889767/FOR5582 ``Modern Foundations of Scattering Amplitudes". RG and TL would like to thank Hadleigh Frost for useful discussions on triangulations of punctured polygons. RG would also like to thank Jacob Bourjaily. JS is grateful to Christoph Bartsch for useful discussions.

\appendix

\section{Refinement of Tree-Level Chambers}\label{sec:app_refinement}
In this appendix we will show that it is necessary to refine tree-level chambers in order to accurately capture the combinatorics of the loop level structure beyond the one-loop fibers. This statement is equivalent to an observation in \cite{He:2023rou} for $n=10$, where it was noticed that the tree-level chambers need to be refined if one studies how the structure of $\Delta^{(2)}({\bf x})$ changes when we vary ${\bf x}$. Our discussion coincides with their results. 

We proceed in the spirit of the discussion outlined in section \ref{sec:half-chambers} by keeping ${\bf x}$ fixed in some tree-level chamber $T=(O,E)$ and studying the refinement of the one-loop fiber $\Delta({\bf x})$ by the lightcones of vertices $q^\pm_{ijk}$. Starting at $n=10$, the geometry of the refined one-loop fibers are no longer specified by the tree-level chambers, and further information is required. A hint that this would happen could already be seen in the decomposition $\Delta^+_{ijk} = \Delta_{\widebar{ij}k} \cup \Delta_{\widebar{i}j\widebar{k}} \cup \Delta_{i\widebar{jk}}$ from equation \eqref{eq:decompose_4}. Let us consider the ten-point example with tree-level chamber $O=(t_{135},t_{159},t_{579}),\, E=(t_{246},t_{2610},t_{6810})$. We are now interested in the vertices of $\Delta_{\widebar{1}3\widebar{5}}=\Delta(x_5,x_6,x_7,x_8,x_9,x_{10},x_1,q^+_{135})$, which is an eight-point one-loop fiber. We know that the vertex structure of an eight-point one-loop fiber is not uniquely determined, as there are four different tree-level chambers. To find which vertices are part of $\Delta_{\widebar{1}3\widebar{5}}$ we ask which triple intersection vertices $q^\pm_{ijk}$ are positively or negatively separated from $q^+_{135}$. 

\begin{figure}
	\centering
	\includegraphics[width=\textwidth]{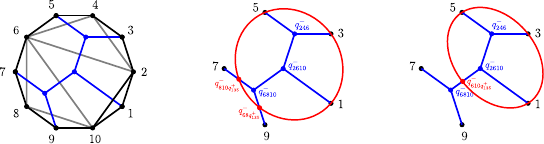}
	\caption{The even triangulation $E=(t_{246},t_{2610},t_{6810})$ (left) and its dual graph in blue. The red ellipse represents the lightcone of $q^+_{135}$ in the cases where $(q^+_{135}-q^-_{6810})^2<0$ (middle) and $(q^+_{135}-q^-_{6810})^2>0$ (right).}
	\label{fig:refinement-example}
\end{figure}

The first important observation is that all odd $q^+_{ijk}$ are always positively separated from one another, and similarly for all even $q^-_{ijk}$. In our example this means that the vertices $q^+_{135}$ and $q^+_{579}$ are part of $\Delta_{\widebar{1}3\widebar{5}}$, which specifies the odd half-chamber structure of this one-loop fiber. What about the even half-chamber? The even tree-level half-chamber of this ten-point $\Delta({\bf x})$ has the structure of a five-point tree-level Feynman diagram, which is depicted in Fig. \ref{fig:refinement-example}. The question about the structure of the even half-chamber of $\Delta_{\widebar{1}3\widebar{5}}$ can then be phrased as a question about which edges of this graph is intersected by the lightcone $\mathcal{N}_{q^+_{135}}$. It turns out that there are two options: 
\begin{itemize}
	\item If $(q^+_{135}-q^-_{6810})^2<0$, then  $\Delta_{\widebar{1}3\widebar{5}}$ has vertices $q^-_{810q^+_{135}}$ and $q^-_{68q^+_{135}}$.
	\item If $(q^+_{135}-q^-_{6810})^2>0$, then  $\Delta_{\widebar{1}3\widebar{5}}$ has vertices $q^-_{6810}$ and $q^-_{610q^+_{135}}$.
\end{itemize}
These possibilities are illustrated in Fig. \ref{fig:refinement-example}.

Whereas the odd/even triangulations of the 10-gon were sufficient to specify the combinatorial structure of $\Delta({\bf x})$, this information is not sufficient to specify the combinatorial structure of $\Delta_{\widebar{1}3\widebar{5}}$ or the fully refined one-loop fiber. Instead, one additional bit of information is needed: namely whether $q^+_{135}$ and $q^-_{6810}$ are spacelike or timelike separated. The surface $(q^+_{135}-q^-_{6810})^2=0$ will \emph{refine} the tree-level chamber into these two cases. 

We thus see that tree-level chambers are refined by the signs of $(q^+_{ijk}-q^-_{abc})^2$. However, not every combination of signs is allowed. In the preceding example, the sign of $(q^+_{135}-q^-_{2610})^2$ is fixed to be negative for all points in this tree-level chamber, and hence there is no further refinement due to this pair of vertices. The reason for this is because the cuts associated to these vertices are `incompatible' in the sense that there is no planar on-shell diagram which realises both cuts at the same time. This can easily be seen by noting that the triangles $t_{135}$ and $t_{2610}$ intersect inside a 10-gon. Of the 25 tree-level chambers at ten points, there are
\begin{itemize}
	\item 10 chambers that do not refine any further, coming from the fact that there are no compatible $q^+_{ijk}$ and $q^-_{abc}$,
	\item 10 chambers that get refined into two smaller chambers, since there is one pair of compatible cuts (like the example above),
	\item 5 chambers which get refined into four smaller chambers, due to the fact that there are two independent pairs of compatible cuts.
\end{itemize}
In total, we thus find that the 25 chambers get refined into 50 chambers, and in each of these refined chambers the structure of the one-loop (half-)chambers will be different. These refined tree-level chambers agree with those found in \cite{He:2023rou}.

\begin{figure}
	\centering
	\includegraphics[width=0.4\textwidth]{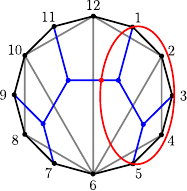}
	\caption{The even triangulation $E=(t_{2\mkern1mu 4 \mkern1mu 6},t_{2\mkern1mu 6\mkern1mu 12},t_{6\mkern1mu 10\mkern1mu 12},t_{6\mkern1mu 8\mkern1mu 10})$ and its dual graph in blue. The lightcone of $q^+_{1\mkern1mu 3\mkern1mu 5}$ is represented by the red ellipse. Although the cut $q^+_{1\mkern1mu 3\mkern1mu 5}$ is compatible with both $q^-_{6\mkern1mu 10\mkern1mu 12}$ and $q^-_{6\mkern1mu 8\mkern1mu 10}$, we see that choosing $(q^+_{1\mkern1mu 3\mkern1mu 5}-q^-_{6\mkern1mu 10\mkern1mu 12})^2>0$ (\textit{i.e.} $q^-_{6\mkern1mu 10\mkern1mu 12}$ is outside the ellipse) automatically implies that also $(q^+_{1\mkern1mu 3\mkern1mu 5}-q^-_{6\mkern1mu 8\mkern1mu 10})^2>0$.}
	\label{fig:refinement-12pt}
\end{figure}

This continues for higher points. Two cuts $q^+_{ijk}$ and $q^-_{abc}$ are compatible if the triangles $t_{ijk}$ and $t_{abc}$ do not intersect, and compatible cuts can be either positively or negatively separated. However, there is one further complication which occurs starting from 12 points. We consider two compatible cuts $q^+_{ijk}$ and $q^-_{abc}$ corresponding to two non-overlapping triangles $t_{ijk}$ and $t_{abc}$. The triangle $t_{abc}$ divides the $n$-gon into three regions with corners $[a,b]=\{a,a+1,\ldots,b\}$, $[b,c]=\{b,b+1,\ldots,c\}$, and $[c,a]=\{c,c+1,\ldots,a\}$. The non-overlapping nature of these triangles means that $\{i,j,k\}$ must be a subset of one of these three index sets, say $[a,b]$. If we consider a tree-level chamber which contains $q^+_{ijk}$ and $q^-_{abc}$, then we can refine this chamber by taking $(q^+_{ijk}-q^-_{abc})^2$ to be either positive or negative. However, if we choose this to be positive, then any triangle in the subpolygons $[b,c]$ and $[c,a]$ must automatically also be positively separated from $q^+_{ijk}$. This is illustrated with an example in Fig. \ref{fig:refinement-12pt}.
\begin{table}
	\centering
	\makebox[\textwidth][c]{
	\begin{tabular}{|c||c|c|c|c|c|c|c|c|c|}
		\hline
		$n$&4&6&8&10&12&14 & 16 & 18 & 20\\
		\hline\hline
		$\#$ tree-level chambers &1&1&4&25&196&1764&17424&184041&2044900\\
		\hline
		$\#$ refined chambers&1&1&4&50&2252&396312& 278970752& 788553663360& 8954861482324480\\
		\hline
	\end{tabular}
	}
	\caption{The number of chambers and refined chambers for $n\leq 20$.}
	\label{tab:number_of_chambers}
\end{table}

We list the number of refined tree-level chambers in Table \ref{tab:number_of_chambers}. We note that these refinements are sufficient to describe the one-loop (half-)chambers, however for higher loop (half-)chambers further refinements might be needed. The logic is similar, to refine an $L$-loop (half-)chamber, we need to consider pairs of compatible vertices of the associated $L$-loop fiber.

\section{One-loop Chambers at Eight Points}
To study the one-loop chambers for eight points, we first fix the null-polygon to be in one of the four tree-level chamber $T=(O,E)$ depicted in Fig. \ref{fig:tree_chambers_8}. As we discussed in section \ref{sec:8pt-half-chamber}, as we vary $y_1$ around $\Delta_T({\bf x})$ we will find exactly eight different structures for both $ \text{skel}^{\downarrow}({\bf x}_{\text{odd}};y_1)$ and $ \text{skel}^{\uparrow}({\bf x}_{\text{even}};y_1)$, which correspond to the one-loop half-chambers which are compatible with $O$ and $E$. This gives a total 64 possible two-loop fibers which we could encounter as we vary $y_1$, of which only 47 are realised. We summarise which two-loop fibers can be found for $O=(t_{137},t_{357}), E=(t_{246},t_{268})$ in Table \ref{tab:8pt-one-loop-chambers}. Using symmetry, we can easily find the possible two-loop fibers for ${\bf x}$ in one of the other three tree-level chambers. This gives us a full classification of the structure of possible two-loop fibers, and hence of all one-loop chambers. We find a total of 157 one-loop chambers at eight-points.
\begin{table}
	\centering
	\makebox[\textwidth][c]{
		\begin{tabular}{c||c|c|c|c|c|c|c|c}
			& \includegraphics[width=1.7cm]{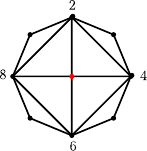} & \includegraphics[width=1.7cm]{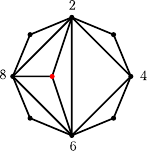} & \includegraphics[width=1.7cm]{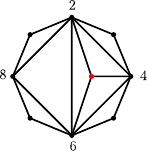} & \includegraphics[width=1.7cm]{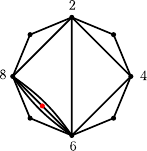} & \includegraphics[width=1.7cm]{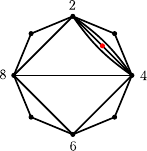} & \includegraphics[width=1.7cm]{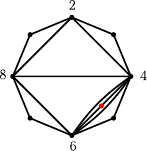} & \includegraphics[width=1.7cm]{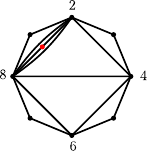} & \includegraphics[width=1.7cm]{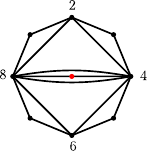} \\
			\hline\hline
			\includegraphics[width=1.5cm]{Figures/eight_point_hc_8} & \color{green}{{\huge \cmark}} & \color{green}{{\huge \cmark}} & \color{green}{{\huge \cmark}} & \color{green}{{\huge \cmark}} & \color{green}{{\huge \cmark}} & \color{green}{{\huge \cmark}} & \color{green}{{\huge \cmark}} & \color{green}{{\huge \cmark}} \\
			\hline
			\includegraphics[width=1.5cm]{Figures/eight_point_hc_7} & \color{green}{{\huge \cmark}} & \color{green}{{\huge \cmark}} & \color{green}{{\huge \cmark}} & \color{green}{{\huge \cmark}} & \color{green}{{\huge \cmark}} & \color{green}{{\huge \cmark}} & \color{red}{{\huge \xmark}} & \color{green}{{\huge \cmark}} \\
			\hline
			\includegraphics[width=1.5cm]{Figures/eight_point_hc_6} & \color{green}{{\huge \cmark}} & \color{green}{{\huge \cmark}} & \color{green}{{\huge \cmark}} & \color{green}{{\huge \cmark}} & \color{green}{{\huge \cmark}} & \color{red}{{\huge \xmark}} & \color{green}{{\huge \cmark}} & \color{green}{{\huge \cmark}} \\
			\hline
			\includegraphics[width=1.5cm]{Figures/eight_point_hc_3} & \color{green}{{\huge \cmark}} & \color{green}{{\huge \cmark}} & \color{green}{{\huge \cmark}} & \color{green}{{\huge \cmark}} & \color{red}{{\huge \xmark}} & \color{green}{{\huge \cmark}} & \color{red}{{\huge \xmark}} & \color{green}{{\huge \cmark}} \\
			\hline
			\includegraphics[width=1.5cm]{Figures/eight_point_hc_1} & \color{green}{{\huge \cmark}} & \color{green}{{\huge \cmark}} & \color{green}{{\huge \cmark}} & \color{red}{{\huge \xmark}} & \color{green}{{\huge \cmark}} & \color{red}{{\huge \xmark}} & \color{green}{{\huge \cmark}} & \color{green}{{\huge \cmark}} \\
			\hline
			\includegraphics[width=1.5cm]{Figures/eight_point_hc_4} & \color{green}{{\huge \cmark}} & \color{green}{{\huge \cmark}} & \color{red}{{\huge \xmark}} & \color{green}{{\huge \cmark}} & \color{red}{{\huge \xmark}} & \color{red}{{\huge \xmark}} & \color{green}{{\huge \cmark}} & \color{red}{{\huge \xmark}} \\
			\hline 
			\includegraphics[width=1.5cm]{Figures/eight_point_hc_2} & \color{green}{{\huge \cmark}} & \color{red}{{\huge \xmark}} & \color{green}{{\huge \cmark}} & \color{red}{{\huge \xmark}} & \color{green}{{\huge \cmark}} & \color{green}{{\huge \cmark}} & \color{red}{{\huge \xmark}} & \color{red}{{\huge \xmark}} \\
			\hline
			\includegraphics[width=1.5cm]{Figures/eight_point_hc_5} & \color{green}{{\huge \cmark}} & \color{green}{{\huge \cmark}} & \color{green}{{\huge \cmark}} & \color{green}{{\huge \cmark}} & \color{green}{{\huge \cmark}} & \color{red}{{\huge \xmark}} & \color{red}{{\huge \xmark}} & \color{red}{{\huge \xmark}}	 
		\end{tabular}
	}
	\caption{Table indicating which one-loop half-chambers intersect for $n=8$ and $T_0^{(o)}=(t_{137},t_{357})$, $T_0^{(e)}=(t_{246},t_{268})$.}
	\label{tab:8pt-one-loop-chambers}
\end{table}

\bibliographystyle{nb}

\bibliography{abjm_chams}
	
\end{document}